\documentclass[%
 reprint,
superscriptaddress,
nofootinbib,
 amsmath,amssymb,
 aps,
pra,
]{revtex4-1}
\usepackage{refcount}
\usepackage{graphicx}
\usepackage{dcolumn}
\usepackage{bm}
\usepackage{color}
\usepackage{mathrsfs}
\usepackage[table]{xcolor}
\usepackage{dsfont}
\usepackage{enumitem}
\setlist[itemize]{leftmargin=0pt}

\begin{document}

\begin{abstract}
Projected squeezed (PS) states are multipartite entangled states generated by unitary spin squeezing, followed by a collective quantum measurement and post-selection. They can lead to an appreciable decrease in the state preparation time of the maximally entangled $N$-qubit Greenberger–Horne–Zeilinger (GHZ) state  when compared to deterministic preparation by unitary transformations in physical systems where spin squeezing can be realized, such as ion, neutral atom, and superconducting qubits. Here we simulate the generation of PS states in non-ideal experimental conditions with relevant decoherence channels. By employing the Kraus operator method, and quantum trajectory method to reduce the computational complexity, we assess the quantum Fisher information and overlap fidelity with an ideal GHZ state. Our findings highlight PS states as useful metrological resources, demonstrating a robustness against environmental effects with increasing qubit number $N$.      
\end{abstract}

\preprint{APS/123-QED}

\title{Robustness of the projected squeezed state protocol}

\author{B. J. Alexander}
 \email{alexander.byronj@gmail.com}
 \affiliation{%
 Department of Physics,
 Stellenbosch University, Matieland 7602, South Africa 
}
\affiliation{%
  Korea Research Institute of Standards and Science,
  Daejeon 34113,
  Korea
}
\author{J. J. Bollinger}
\affiliation{
National Institute of Standards and Technology, Boulder, Colorado 80305, USA}

\author{M. S. Tame}%
\affiliation{%
 Department of Physics,
 Stellenbosch University, Matieland 7602, South Africa 
}

\date{\today}
\maketitle

\section{Introduction}
When implementing quantum protocols in realistic physical systems, the quantum resource state is invariably influenced by the system's interaction with its surrounding environment~\cite{schlosshauer2019quantum, nielsen2002quantum}. Thus, to ensure the reliability of the results, it is necessary to account for the impact of the coupling between the system and the environment. Emerging quantum technologies rely principally on quantum phenomena such as entanglement to surpass classical limitations~\cite{horodecki2009quantum}.~It is therefore an essential task to develop robust protocols for producing and controlling classes of highly entangled quantum states in the presence of experimentally relevant noise sources. Typically, the entanglement between the system and the environment leads to a loss of quantum coherence, known as quantum decoherence.

An important entangled state in quantum information protocols is the maximally entangled $N$-qubit Greenberger-Horne-Zeilinger (GHZ) state~\cite{greenberger1989going}, given as
\begin{align}
|GHZ\rangle := \frac{|0\rangle^{\otimes N} + |1\rangle^{\otimes N} }{\sqrt{2}},\label{GHZeq}
\end{align}
where $|0\rangle$ and $|1\rangle$ are the computational basis states of a single qubit. The symbol $\otimes N$ denotes the tensor product of the individual states of the composite $N$-qubit system. 
This is a versatile entangled state with applications in various areas of quantum information processing~\cite{zheng2000efficient}, including quantum metrology~\cite{schaffry2010quantum, pezze2018quantum}, quantum cryptography~\cite{broadbent2009ghz, zhu2018semi}, and quantum communication~\cite{gao2005deterministic,jin2006three}. Recently, this state has been experimentally realized deterministically with superconducting qubits by engineering a one-axis twisting Hamiltonian \cite{song2019generation}. Experimental realizations have also been reported with trapped ions~\cite{leibfried2005creation,monz201114} and Rydberg atoms~\cite{omran2019generation}. Further theoretical proposals that utilize spin squeezing to generate highly entangled GHZ-like states include ultra-cold atoms~\cite{plodzien2020producing}, phonon-spin ensembles~\cite{xia2016generating}, Bose-Einstein condensates~\cite{gross2012spin}, and trapped ions~\cite{foss2013nonequilibrium}. For examples of measurement-based schemes for generating GHZ states, see Refs.~\cite{helmer2009measurement} and \cite{bishop2009proposal}.

 In this study we build upon our previous work that introduced a method for generating, from an initial separable state, a class of GHZ-like states that exhibit genuine multipartite entanglement~\cite{toth2005entanglement}, known as projected squeezed (PS) states (see Alexander et al. in Ref.~\cite{alexander2020generating}). Essential steps of the PS state generation protocol include strong spin squeezing to generate a spread or wrap of the composite spin probability distribution around the Bloch sphere, followed by a collective quantum state measurement of a component of the composite spin. Based on the measured value of the spin component, the resulting spin state can approximate a superposition of two different coherent spin states pointing in opposite directions, resembling the GHZ state given in Eq.~(\ref{GHZeq}). Atomic spins in optical cavities provide an example of a system where the required collective spin measurement can be implemented~\cite{schleier2010squeezing, chen2011conditional, zhang2012collective, hosten2016measurement}.

We explore the generation and properties of PS states within the context of decoherence, focusing on their advantages over deterministic preparation methods for achieving the GHZ state. PS states, generated through unitary spin squeezing and post-selection, offer a significant decrease in state preparation time compared to unitary evolution for systems with large $N$ or high decoherence. The class of PS states described in Ref.~\cite{alexander2020generating} was shown to yield favorable overlap fidelity with the GHZ state of approximately $\mathcal{F} \gtrsim 0.90$ for $N \gtrsim 30$, with an upward monotonic trend observed for increasing system size $N$.

Our investigation involves simulating the generation of PS states under non-ideal experimental conditions. Utilizing the Kraus operator method and the quantum trajectory method to reduce computational complexity, we evaluate the quantum Fisher information and overlap fidelity of PS states with an ideal GHZ state. To analyze the impact of decoherence during the PS state generation, we consider experimentally relevant decoherence channels~\cite{carlo2004simulating, foss2013nonequilibrium}. Our objective is to demonstrate the superior performance of the PS state protocol compared to deterministically generated macroscopic-superposition states (MSSs) in the presence of decoherence, particularly as the system size $N$ increases. It is worth noting that the GHZ state can be derived from the MSS state using a sequence of local unitary operations~\cite{pezze2018quantum, pezze2019heralded}.

The outcomes of our study emphasize the utility of PS states as valuable metrological resources. We demonstrate their scalability and robustness against environmental effects, showcasing their potential for reliable quantum information processing. By providing a comprehensive analysis of the implications and performance of PS states under decoherence, our findings contribute to advancing the understanding and practical application of these states.

In Section II, we provide a brief overview of the requisite concepts in quantum information theory. In particular we define the important metrological quantities of the fidelity of quantum states, and the quantum Fisher information (QFI). These definitions are necessary for subsequent discussions and analyses.

Section III presents a detailed step-by-step description of the modified PS state protocol, along with an elaboration on the unitary MSS protocol, which is used as a benchmark for comparison. We aim to provide a clear understanding of the procedures involved and their relevance to our research objectives. Additionally, we contextualize the motivation behind our research, highlighting the modification of the original PS state protocol for the purpose of yielding superior resource states for quantum sensing.

In Section IV, we analyze the computational complexity of the ideal decoherence-free PS state protocol in comparison to the protocol incorporating relevant decoherence channels. Furthermore, we describe how the reduction in computational complexity serves as a compelling reason for adopting the quantum trajectory method.

In Section V, we describe and motivate the quantum trajectory method, used to simulate open quantum systems. Here we highlight the advantages and rationale of utilizing this method, underscoring its relevance in efficiently evaluating central metrological quantities.

To facilitate the practical implementation, Section VI presents the numerical methods utilized for executing the Kraus operator sum method and quantum trajectory method. We provide a detailed description, considering relevant experimental parameters, with a particular focus on their applicability within a trapped-ion experimental setup. This section provides a comprehensive comparative analysis of the efficiency between the PS state protocol, accounting for decoherence effects, and the unitary MSS generating protocol, which serves as a benchmark for reference. By illustrating these results, we highlight the influence of decoherence on the performance of the PS state protocol, further emphasizing the significance of our modified approach.

Section VII offers a discussion and summary of the key findings presented in the study.

\begin{figure*}[tbp]
\hspace*{-0.0cm}
  \centering
  \begin{tabular}{ccc}
\includegraphics[width=172mm,scale=0.99]{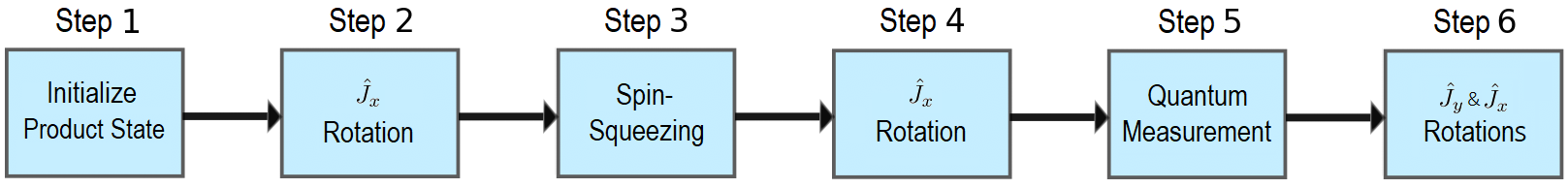}&
  \end{tabular}
  \caption{An overview of the PS state protocol. See main text for details.}
  \label{A}
\end{figure*}

\section{Preliminaries} 
To quantify the notion of `closeness' between two quantum states, we consider the overlap fidelity. Although not a metric, as it does not fulfill the triangle inequality, it serves as a useful and widely used quantum state comparative tool. The overlap fidelity, similar to the classical Bhattacharyya distance measure~\cite{derpanis2008bhattacharyya}, operates as a statistical distance for quantifying the similarity between two probability distributions. Specifically, the comparison focuses on the marginal probability distributions (or probability amplitudes) of two quantum states~\cite{nielsen2002quantum}, rather than their joint distribution. Given density operators $\rho$ and $\sigma$, the fidelity is defined as
\begin{align}
    \mathcal{F}(\rho, \sigma) :=  \bigg(\text{tr}  \sqrt{ \sqrt{\rho }\sigma\sqrt{\rho }}\bigg)^2, \label{OV}
    \end{align}
    where $\sqrt{\cdot}$ denotes the matrix square root. Given pure states $\rho = |\psi\rangle \langle \psi|$ and $\sigma = |\phi\rangle \langle \phi|$, this simplifies to 
    \begin{align}
    \mathcal{F}(\rho, \sigma) =  \big|\langle \phi | \psi \rangle \big|^2. \label{OV2}
    \end{align} 
 The fidelity is a useful and well-established quantity for measuring the closeness of two quantum states~\cite{nielsen2002quantum}. Hereafter, the context of the fidelity will mainly be with respect to the PS and GHZ states, which we will simply denote as $\mathcal{F}$.

An entanglement witness is an operator that is used to distinguish between multipartite entangled states and separable states. An Hermitian operator $\hat{W}$ is an entanglement witness if it yields non-negative expectation values for separable states, and negative values for some entangled states,
\begin{equation}
  \text{Tr}(\hat{W}\rho)~~\begin{cases}
    ~~\geq 0, & \text{for all separable states,}\\
    ~~< 0, & \text{for some entangled states.}
  \end{cases}
\end{equation}
In Ref.~\cite{toth2005entanglement}, a detailed discussion is provided on constructing entanglement witnesses, with a particular focus on states that exhibit high fidelity with the GHZ state. 

An example of a projector witness that detects genuine multipartite entanglement (GME) of states close to the GHZ state \cite{toth2005entanglement}, reads as
\begin{equation}
\hat{W}_{\text{GHZ}}:= \frac{1}{2}\hat{\mathds{1}}-|GHZ\rangle\langle GHZ|.
\label{GHZwit}
\end{equation}
 This entanglement witness provides a sufficient condition for GME based on the state's fidelity with the GHZ state, specifically implying that $\mathcal{F} > 1/2$ indicates the presence of GME. 
Furthermore, experimental implementations of the concept of entanglement witnesses have been reported~\cite{bourennane2004experimental, zhu2022experimental}.

The quantum Fisher information (QFI)~\cite{petz2011introduction, paris2009quantum}, which is the quantum analogue of the classical Fisher information~\cite{helstrom1969quantum, braunstein1994statistical}, is an important quantity in quantum metrology~\cite{lu2015robust, arvidsson2020quantum}. Furthermore, there exist established theoretical benchmarks that connect the multipartite entanglement of a quantum state with its QFI~\cite{toth2012multipartite}. More commonly, the QFI is used to quantify the amount of information about a parameter encoded in a state via a unitary evolution. Consider some unitary transformation given by 
\begin{align}
    \hat{U}(\theta'):=\text{exp}\big(-i \hat{H} \theta' t/ \hbar\big),
    \label{UH}
\end{align}
where $\hat{H}$ denotes some Hamiltonian operator, while $\theta'$ denotes the unitary phase parameter. For a simplified representation we consider the phase parameter $\theta := \theta' t/ \hbar$, which depends on the evolution time and the system properties, yielding $\hat{U}(\theta) = \text{exp}(-i \hat{H} \theta)$ from Eq.~(\ref{UH}).
The QFI with respect to $\hat{H}$, for some state $\rho$, is given by (see Refs.~\cite{helstrom1969quantum, braunstein1994statistical, paris2009quantum})
\begin{align}
    \mathcal{Q}(\rho, \hat{H}):= 2\sum_{\substack{j,k=1 \\ \lambda_{j}+\lambda_{k} \neq 0}}\frac{(\lambda_j-\lambda_k)^2}{\lambda_j+\lambda_k}\big|\langle j | \hat{H}|k\rangle\big|^2,
    \label{QFI}
\end{align}
where $\lambda_i$ and $|i\rangle$ are the eigenvalues and eigenvectors of $\rho$, respectively.

The QFI exhibits convexity \cite{toth2014quantum} and therefore, for the mixed state $\rho = p\rho_{1} + (1-p)\rho_{2}$, the following inequality holds
\begin{align}
\mathcal{Q}(\rho, \hat{H}) \leq p\mathcal{Q}(\rho_{1}, \hat{H})+(1-p)\mathcal{Q}(\rho_2, \hat{H}).
\label{conv}
\end{align}
Furthermore, it is known that the QFI of a state $\rho$ with respect to some Hamiltonian $\hat{H}$, satisfies the inequality 
\begin{align}
    \mathcal{Q}(\rho, \hat{H}) \leq 4(\Delta H)^2
    \label{EQ},
\end{align} where $(\Delta H)^2 = \langle (\Delta \hat{H})^2 \rangle = \langle \hat{H}^2 \rangle - \langle \hat{H} \rangle^2$ represents the uncertainty (or variance) in the measurement of the system's energy, and where equality holds for pure states~\cite{paris2009quantum}.

Quantum states with favorable QFI are suitable resource states for the canonical phase estimation task, which essentially seeks to estimate, to highest precision, the encoded phase $0<\theta \ll 1$ of a state $\rho$ after some unitary transformation, that is $\hat{U}(\theta):\rho \mapsto \hat{U}(\theta)\rho \hat{U}^{\dagger}(\theta)$, with $\hat{U}(\theta)$ defined by Eq.~(\ref{UH}). The QFI constrains the achievable estimation precision of $\theta$, by setting a lower-bound on the estimator variance, which is defined by
\begin{align}
    (\Delta\theta)^2:=\frac{(\Delta A)^2}{\big|\partial_{\theta} \langle \hat{A} \rangle \big|^2},
    \label{DD}
\end{align}
where $\langle \hat{A} \rangle$ and  $(\Delta A)^2:=\langle \hat{A}^2 \rangle-\langle \hat{A} \rangle^2$ denote the expectation value and variance of a chosen measurement operator $\hat{A}$, respectively. The phase estimation precision, for any single-shot measurement, is bound by the well-known quantum Cram\'{e}r-Rao bound~\cite{helstrom1969quantum, braunstein1994statistical, paris2009quantum}, which reads as 
\begin{align}
    (\Delta \theta)^2 \geq \frac{1}{\mathcal{Q}(\rho, \hat{H})}.
     \label{CR}
\end{align}

\section{The ideal PS state protocol}
In Ref.~\cite{alexander2020generating} the ideal decoherence-free PS state protocol is introduced and described in the subspace spanned by the symmetric Dicke basis states~\cite{dicke1954coherence}. This subspace is no longer suitable when including decoherence channels, since as a consequence, the state can migrate out of the subspace.  Our approach involves modifying the original set of measurement operators introduced in Ref.~\cite{alexander2020generating}. This modification enables the PS state protocol to be well-defined in the total Hilbert space, even with the inclusion of decoherence. Additionally, we reduce the magnitude of spin squeezing and incorporate an extra collective Pauli-X rotation in the final step. 

For an $N$-qubit system, the collective Pauli-X rotation operator is defined by \begin{align}
\hat{J_{x}}:= \frac{1}{2}\sum_{i=1}^{N}\hat{\sigma}_{i}^{x},\end{align}    
where ${\hat{\sigma}}_{i}^{x}$ denotes the Pauli-X operator associated with the $i$-th qubit. The collective Pauli-Y and Pauli-Z rotation operators are similarly defined, and denoted by $\hat{J}_{y}$ and $\hat{J}_{z}$, respectively. The modified protocol we present in this work achieves superior fidelity and QFI results for all $N$, while requiring reduced spin squeezing compared to the findings reported in Ref.~\cite{alexander2020generating}. These results are described in detail in subsequent sections.    
\begin{figure*}[tbp]
\hspace*{-0.25cm}
  \centering
  \begin{tabular}{ccc}
\includegraphics[width=178.5mm,scale=0.99]{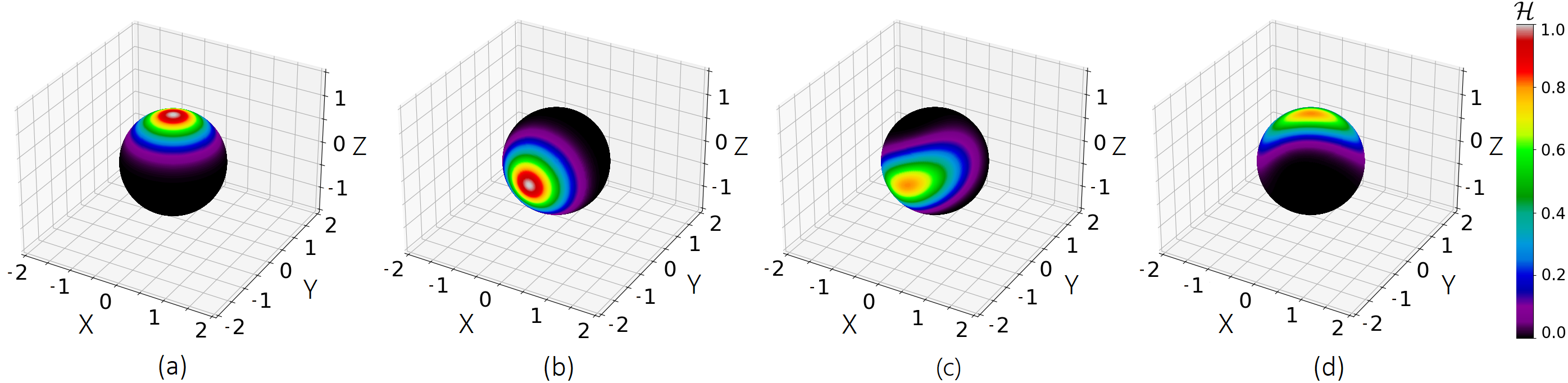}
  \end{tabular}
  \caption{The Husimi quasiprobability phase space distributions $\mathcal{H}$ of the pre-measurement states described by steps 1 - 4 of the PS state protocol for $N=10$. (a) Initialize to all spin up state $|\psi\rangle = |\uparrow \rangle \otimes \cdot \cdot \cdot \otimes |\uparrow \rangle$, (b) Step 2 - form the coherent spin state by a  collective $\hat{J}_x$ rotation by $\frac{\pi}{2}$ rad, (c) Step 3 - the coherent spin state undergoes spin squeezing by $\chi t = 0.15$, (d) Step 4 - the spin-squeezed state undergoes a collective $\hat{J}_x$ rotation by~$-\frac{\pi}{2}$ rad.}
  \label{B}
\end{figure*}

The dimension of the total Hilbert space scales exponentially with the system size
$(\sim 2^{N} )$, while the dimension of the symmetric subspace spanned by the Dicke-basis has linear
scaling $(\sim N+1)$. This increase in computational complexity restricts our ability to simulate larger system sizes within a reasonable time frame. To overcome these computational limitations, we utilize the quantum trajectory method~\cite{gardiner1992wave, breuer1995stochastic, carlo2004simulating}, which allows us to model the protocol as a quantum state vector, as opposed to the more computationally expensive density operator. As a result, we are able to evaluate the QFI using the simpler and computationally cheaper Eq.~(\ref{EQ}) for pure states, as opposed to Eq.~(\ref{QFI}) which requires diagonalizing the density operator, the computational complexity of which increases substantially as a function of the state space dimension. Based on Eqs.~(\ref{conv}) and (\ref{CR}), the average QFI obtained from the quantum trajectories establishes a lower bound on the precision $(\Delta \theta)^2$. The time required to computationally simulate the density operator evolution is only practical for low system sizes of the order $N \lesssim 10$. For larger $N$, we used the quantum trajectory method to simulate the preparation of PS states while taking into account experimentally relevant decoherence channels~\cite{foss2013nonequilibrium}. 

The PS states are a class of highly entangled $N$-qubit states~\cite{alexander2020generating}, generated by a particular sequence of collective rotations, unitary spin squeezing~\cite{kitagawa1993squeezed} and collective quantum measurement with post-selection. The protocol for producing PS states, for arbitrary $N$, can be summarized by the following sequential steps (see Fig.~\ref{A}):
\begin{itemize}
    \item[] {\bf Step 1}: Initialize a pure $N$-qubit all spin-up state,  \begin{align}|\psi (0)\rangle = \underbrace{| \uparrow \rangle \otimes \cdot \cdot \cdot \otimes |\uparrow \rangle}_\text{$N$-qubits},\label{Init}
    \end{align}
    where $|\uparrow \rangle:=|0\rangle$ and $|\downarrow\rangle:=|1\rangle$.
     \item[] {\bf Step 2}: To form the coherent spin state, denoted by $|CS\rangle$, execute a  $\pi/2$ collective $\hat{J}_{x}$ rotation 
    on the all-spin up state,
    \begin{align}
    |\psi(0)\rangle \mapsto \text{exp}\bigg(-i\frac{\pi}{2}\hat{J_x}\bigg)|\psi(0)\rangle =: |CS\rangle.\end{align}

    \item[] {\bf Step 3}: The coherent spin state then undergoes unitary spin squeezing characterized by the one-axis twisting operator 
    \begin{align}
         \hat{U}_{Sq}(\chi t) := \text{exp}\big(-i \chi t \hat{J}^2_{z}\big),\label{Sq}
    \end{align}
    where $\chi t$ denotes the strength of the interaction. The one-axis twisting operator reduces the spin uncertainty of the state along one spin axis, at the expense of increasing the uncertainty along an orthogonal spin axis~\cite{uys2012toward}. We squeeze with sufficient magnitude to yield states with a positive probability wrap about the Bloch sphere (see the Husimi quasiprobability distribution in Fig.~\ref{B}(c)); required for the later quantum measurement described in step 5. The Husimi distribution of a state $|\psi\rangle$ (see Refs.~\cite{uys2012toward, husimi1940some}), is the modulus squared of the projection of the state onto a rotated coherent spin state, that is $\mathcal{H}:=\big|\langle\psi|\text{exp}(-i \phi \hat{J}_{z}) \text{exp}(-i \theta \hat{J}_x)| CS\rangle\big|^2$, where $\theta$ and $\phi$ are respectively the polar and azimuthal angles. This is the chosen phase space quasiprobability distribution as it is both intuitive for visualizing symmetric Dicke states~\cite{dicke1954coherence}, and consistent with the original study~\cite{alexander2020generating}. In principle, any quasiprobability distribution could be used. 
    \item[] {\bf Step 4}:  The squeezed coherent spin state then undergoes a -$\pi/2$ collective $\hat{J}_{x}$ rotation, 
    \begin{align}
    \hspace{0.85cm}
        \hat{U}_{Sq}(\chi t)|CS\rangle \mapsto \text{exp}\bigg(i\frac{\pi}{2}\hat{J_x}\bigg) \hat{U}_{Sq}(\chi t)|CS\rangle.
    \end{align}   
     See Fig.~\ref{B} for Husimi distributions of the state in the pre-measurement steps~1-4.
    \item[] {\bf Step 5}: We now perform a quantum measurement described by a positive operator-valued measure (POVM) (see Ref.~\cite{jacobs2014quantum}). The quantum measurement is characterized by a set of linear positive semi-definite operators $\{\hat{A}_c\}_{c\in \mathbb{R}}$ on the complex Hilbert space. These operators can be expressed as 
    \begin{multline}
    \hspace{-0.0cm}
    \hat{A}_c := \sum_{m=0}^{N}
    \sqrt{\text{Pr}\left(\frac{N}{2}-m\Big|c\right)}\sum_{\underline{M}} \\
    \hspace{0.8cm}
    \times \underbrace{\big| \uparrow \cdots \downarrow_{j_1}\cdots \downarrow_{j_m} \cdots\uparrow \big\rangle}_\text{$N$-qubits} \big\langle \uparrow \cdots \downarrow_{j_1}\cdots \downarrow_{j_m}\cdots\uparrow \big|.
    \label{Ac}
    \end{multline}
  In this expression, $c \in \mathbb{R}$ represents the measurement outcome, which corresponds to a set with a continuum cardinality. The summation $\sum\limits_{\underline{M}}\cdot$ accounts for all binary permutations $\underline{M}$ of length $N$ with $m$ qubits in the spin-down state. 
    The projector weightings of Eq.~(\ref{Ac}) are characterized by Gaussian probability distributions
    \begin{align}
    \text{Pr}\big(x|c\big):= \frac{1}{\sqrt{2 \pi \sigma^2}}\exp \bigg[-\frac{(x-c)^2}{2 \sigma^2} \bigg],\label{Pr}    
    \end{align}
where $\sigma^2\geq0$ modulates the spread of the probability amplitudes over the projectors that define the measurement. This is a modification of the set of measurement operators described in the original study~\cite{alexander2020generating}. When considering a modification, there is a level of freedom in choosing how to distribute the projector weightings given by Eq.~(\ref{Pr}) without compromising the required completeness condition, i.e., $\int \hat{A}_c \hat{A}_c^{\dagger} dc = \hat{\mathds{1}}$. To this end, Eq.~(\ref{Ac}) is a natural extension to the total Hilbert space of the measurement operators described in Ref.~\cite{alexander2020generating}. 

Appendix A provides the equivalence of the original and modified measurement operators in the decoherence-free setting, where they are both well-defined in the symmetric Dicke subspace.

Atomic ensembles in optical cavities provide an experimental system where unitary single-axis twisting and collective state measurement described by Eqs.~(\ref{Ac}) and (\ref{Pr}) has been demonstrated with laser-cooled, dilute atomic gasses~\cite{schleier2010squeezing, chen2011conditional,zhang2012collective,hosten2016measurement}. Optical cavity measurement techniques can in principle also be used with ensembles of trapped ions. In trapped-ion systems, another potential method for executing the POVM described by Eqs.~(\ref{Ac}) and (\ref{Pr}) is to do a state-dependent excitation of the ion motion using the optical dipole force~\cite{metcalf2003laser}. The optical dipole force is a spin-dependent force which is derived from the gradient of the potential of an inhomogeneous electromagnetic field. For 2-D ion crystals in Penning traps, the optical dipole force is arranged to act in the z-direction (see Refs.~\cite{bohnet2016quantum} and \cite{britton2012engineered}).  If resonant with the center-of-mass mode, this excites the center-of-mass mode with a strength proportional to $N/2 – m$. The image current induced in the ion trap electrodes is expected to be proportional to the projection quantum number $N/2 – m$, as opposed to the individual ion state. Measuring the image current induced by the ion crystal motion can therefore in principle execute the measurement given by Eq.~(\ref{Ac}). A given measurement of the image current would then give a fixed $c$ value, while $\sigma$ corresponds to the systematic precision with which the measurement projects out states about the value $c$.

\begin{figure*}
\hspace*{-0.25cm}
  \centering
  \begin{tabular}{ccc}
\includegraphics[width=178.5mm,scale=0.99]{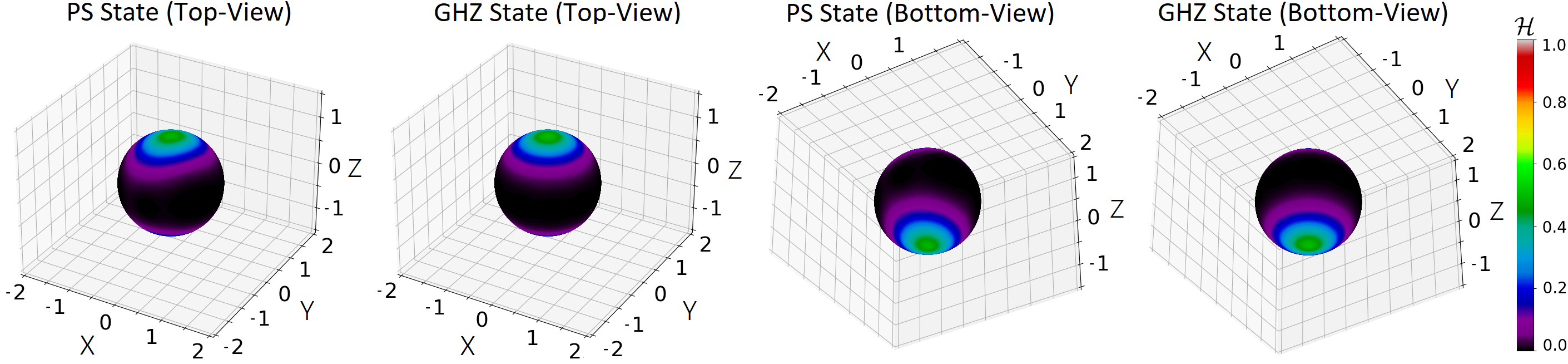}
  \end{tabular}
  \caption{The PS state and GHZ state Husimi distributions $\mathcal{H}$ for $N=10$.}
  \label{C}
\end{figure*}
The post-measurement state reads as 
\begin{align}
    \frac{\hat{A}_c \rho  \hat{A}_{c}^{\dagger}}{\text{Tr}\big[\hat{A}_c^{\dagger}\hat{A}_c\rho\big]},
    \label{qm}
\end{align}
for measurement outcome $c$, which occurs with probability $\text{Tr}\big[\hat{A}_c^{\dagger}\hat{A}_c\rho\big]$. A quantum measurement is an inherently stochastic process, and as such, a desired measurement outcome requires post-selection. Although the term projected squeezed state is used in this work and in the original study~\cite{alexander2020generating}, a more precise description of the quantum measurement outlined in Eqs.~(\ref{Ac}) and (\ref{Pr}) is as a POVM. This is due to the measurement's non-orthogonality for different outcomes, continuous spectrum, and fulfillment of the completeness condition. 

In the original study, the optimal measurement outcome was reported as $c \approx 0$. However, with the implementation of the modified protocol (see step 6), we have found the optimal measurement outcome to be $c \approx -2.5$. The computational analysis suggests that the optimal post-selected measurement outcome could depend on $N$, but for the limited system size interval considered here, we find a common optimal measurement outcome of $c \approx -2.5$. In Ref. \cite{alexander2020generating}, which describes an ideal protocol in the Dicke subspace (without the inclusion of decoherence channels), the optimal measurement outcome of $c \approx 0$ has a clearer geometrical interpretation when represented as a Husimi distribution than the modified protocol introduced in this study. This clarity arises from the Dicke subspace being spanned by symmetric states, with a quoted total spin-squeezing of $\chi t \approx 0.4$ resulting in a probability wrap about the Bloch sphere. The quoted optimal post-selected outcome of $c=0$ can be represented as the projection about the equator of the multi-qubit Bloch sphere, where the projection probability amplitudes peak around what can be considered as the central basis state, the Dicke state $|\frac{N}{2},0\rangle$.
However, with the inclusion of decoherence channels, the modified protocol, now presented in the full Hilbert space, essentially describes a new protocol, with an additional collective $\hat{J}_{x}$ rotation allowing substantially reduced spin-squeezing requirements. This reduction in spin-squeezing results in a shifting of the optimal measurement outcome. The optimal measurement outcome is thus a function of the magnitude of total spin-squeezing. The optimal post-measurement outcome of $c = -2.5$ has a less intuitive geometrical interpretation but should rather be regarded as a continuous control parameter chosen to maximize the overlap fidelity with the GHZ state, while minimizing the required duration of the one-axis squeezing. A comprehensive motivation of this optimal measurement will be provided. 

Furthermore, the modified measurement operator given in Eq.~(\ref{Ac}) becomes necessary with the inclusion of decoherence, which generates states lying outside the Dicke manifold of symmetric spin states. In general, the original measurement operator given in Ref.~\cite{alexander2020generating} will not preserve the relative phase of superposition states in the general $2^N$ dimensional Hilbert space spanned by $N$ spins. This modification makes the measurements more representative of the way they would be performed in an experiment, with the relative phases preserved. 

    \item[] {\bf Step 6}: Finally, we sequentially perform $\hat{J}_y$ and $\hat{J}_x$ rotations using the computationally determined optimal angles of $\pi/2$ rad and $5.6$ rad, respectively. These rotation angles are observed to be optimal for arbitrary$~N$. 

\end{itemize}

The fidelity $\mathcal{F}$ serves as a measure of how closely the PS state approximates the GHZ state (see Fig.~\ref{C} for the corresponding Husimi distributions for $N=10$). The final $\hat{J}_{x}$ rotation was not included in the original study~\cite{alexander2020generating}, but on further analysis it is shown to consistently yield superior GHZ overlap fidelity for varied system sizes.  This process allows us to generate a PS state with maximal fidelity $\mathcal{F}$. 
Through the utilization of computational methods, we have determined the optimal values of $c \in \mathbb{R}$, $\sigma^2 \geq 0$ (as defined in Eq.~(\ref{Pr})), collective rotation angles, and total squeezing magnitude $\chi t \geq 0$ that maximize the fidelity $\mathcal{F}$. Moreover, to simulate and analyze the impact of decoherence during the PS state protocol, we incorporate experimentally relevant decoherence channels~\cite{carlo2004simulating, foss2013nonequilibrium}. 

Our goal is to demonstrate that with relevant experimental decoherence, for increasing system size $N$, the PS state protocol yields increasingly superior fidelity results when compared to the MSS generating protocol. The MSSs are a class of entangled states deterministically generated by spin squeezing the coherent spin state (see Ref.~\cite{foss2013nonequilibrium}). For even $N$, a subsequent $\hat{J}_{x}$ rotation by $\pi/2$ of the MSS, followed by a local phase gate  $S := \scriptsize\begin{pmatrix}
1 & 0 \\
0 & i 
\end{pmatrix}$ on the $N$-th qubit, yields the GHZ state given by Eq.~(\ref{GHZeq}). The MSS is therefore local unitary (LU) equivalent to the GHZ state, thus yielding entangled quantum states with GHZ overlap fidelity of unity, which we denote by $\mathcal{F}_{MSS}=1$.

We will show that the PS state protocol is a robust scheme for generating highly-entangled quantum states, and a suitable alternative to the deterministic unitary MSS scheme for generating GHZ states. In this way, the MSS scheme serves as a benchmark for comparing the efficacy of the PS state protocol. The PS states have a significant advantage over this benchmark in terms of their total spin squeezing requirements, which are substantially reduced. This implies that, in principle, the PS state protocol should perform better in the presence of decoherence, since it is exposed for a shorter duration. Upon comparing the PS state protocol with the MSS scheme, we confirm that this is indeed the case.  

The inclusion of the quantum measurement described in step 5 introduces non-unitary dynamics, which implies probabilistic non-deterministic post-measurement states. To this end, we must consider the success rates of post-selecting desired PS states. We look at this in detail.  

\begin{figure}[b]
\hspace*{-0.0cm}
  \centering
  \begin{tabular}{ccc} \includegraphics[width=71mm,scale=0.99]{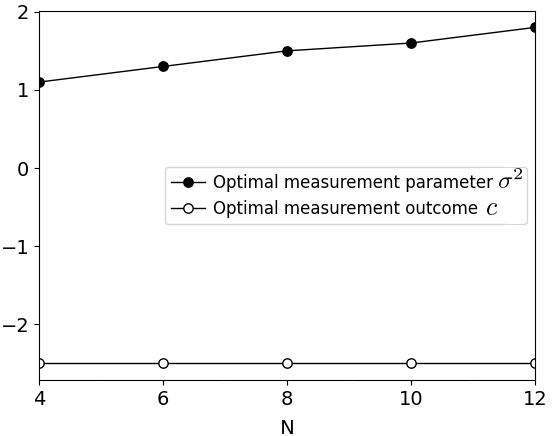}
  \end{tabular}
  \caption{The identified optimal measurement outcome $c$ and measurement parameter $\sigma^2$ for generating PS states with maximal $\mathcal{F}$, for $N \in \{4, 6, 8, 10, 12\}.$}
  \label{var}
\end{figure}

For $N=10$, Fig.~\ref{C} compares Husimi distributions of an optimized PS state with the target GHZ state. The identified optimal total evolution of $\chi t \approx 0.15$, optimal post-selected measurement outcome of $c \approx -2.5$, and optimal measurement parameter of $\sigma^2 \approx 1$ (see Eq.~(\ref{Pr}))\footnote{In the projector probability distribution of Eq.~(\ref{Pr}), $\sigma^2 \approx 1$ corresponds to an ability to measure the number of qubits in the spin-down state at the level of 1 or 2 qubits.} generates a PS state with a favorable GHZ overlap fidelity of $\mathcal{F} > 0.99$. The probability lobes on the top and bottom of the Bloch sphere are a characteristic feature of highly entangled GHZ-like states. The identified evolution of $\chi t = 0.15$ is notably shorter than the $\chi t = \pi/2$ required by unitary single-axis twisting to generate MSSs, by more than an order of magnitude. 

 In the decoherence-free setting, the computational results indicate that for $N\in\{4, 6, 8, 10, 12 \}$, the evolution of $\chi t = 0.15$, and optimal measurement of $c = -2.5$ yield PS states with $\mathcal{F} \gtrsim  0.98$. While this optimal measurement outcome is not necessarily unique for a given $N$, it is beneficial to fix some optimal measurement for the purpose of numerical optimization. We consider a measurement optimal if it yields $\mathcal{F}>0.98$ in the decoherence-free setting. In Fig.~\ref{var} we present the identified optimal measurement parameter $\sigma^2$ that characterizes the optimal measurement operator $\hat{A}_{c}$ defined by Eq.~(\ref{Ac}). The optimal $\sigma^2$ depends on the chosen post-selected measurement outcome $c$ and system size $N$. For increasing $N$, a clear upward monotonic trend is observed.     

\begin{figure}
\hspace*{-0.15cm}
  \centering
  \begin{tabular}{ccc}
\includegraphics[width=83.75mm,scale=0.99]{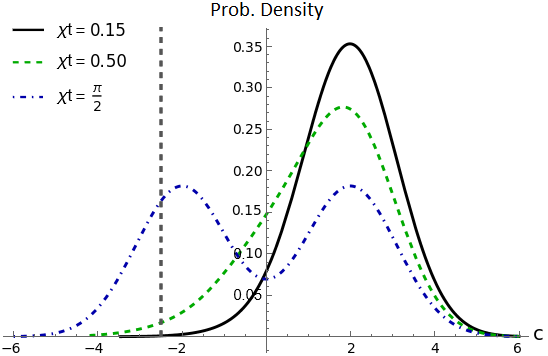}
  \end{tabular}
  \caption{The post-selection probability density function (PDF) for $N=4$ and $\sigma^2 = 1.1$, with varied spin squeezing of $\chi t \in \{0.15, 0.5, \frac{\pi}{2} \}$. The dashed vertical at $c=-2.5$ denotes the identified optimal measurement.}
  \label{PDF}
\end{figure}

\begin{figure}
\hspace*{-0.19cm}
  \centering
  \begin{tabular}{ccc}
\includegraphics[width=84.75mm,scale=0.99]{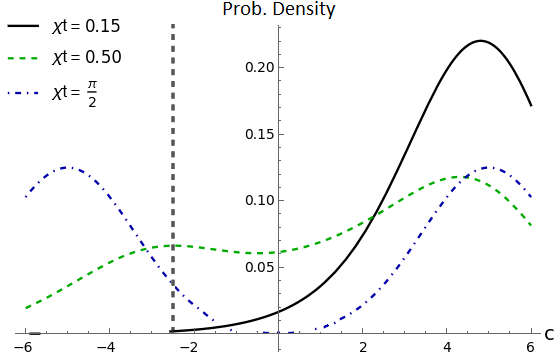}
  \end{tabular}
  \caption{The post-selection probability density function (PDF) for $N=10$ and $\sigma^2 = 1.6$, with varied spin squeezing of $\chi t \in \{0.15, 0.5, \frac{\pi}{2} \}$. The dashed vertical at $c=-2.5$ denotes the identified optimal measurement.}
  \label{PDF10}
\end{figure}

For $N=4$ and $N=10$, the post-selection probability density functions (PDFs) for measurements $c \in [-6, 6]$ and total evolution of $\chi t \in \{0.15, 0.5, \pi/2 \}$ are presented in Fig.~{\ref{PDF}} and Fig.~{\ref{PDF10}}, respectively. In Fig.~{\ref{PDF}} it is apparent that an increase from $\chi t = 0.15$ to $\chi t = \pi/2$ results in an increased probability of obtaining outcomes about the identified optimal measurement $c=-2.5$. The increased squeezing distributes the probability amplitudes such that the post-measurement state has an increased probability of producing high fidelity PS states. With total evolution of $\chi t = 0.15$, Fig.~\ref{B}(c), Fig.~\ref{PDF}, and Fig.~\ref{PDF10} demonstrate that we achieve sufficient squeezing to induce wrapping around the Bloch sphere for a positive measurement projection about the optimal outcome $c=-2.5$. As a result, with $\chi t = 0.15$, the optimal post-selection probabilities are distributed asymmetrically about $c =-2.5$. Note that even though the PDFs are peaked at $c \approx 2$ and $c \approx 4.5$, for $N=4$ and $N=10$, respectively, projections about these values do not lead to desirable states. An increase in total evolution results in a more desirable probability profile, suggesting a trade-off, since an increase also requires more time and thus exposes the quantum state to further decoherence. Given this trade-off, we consider the fidelity as the primary indicator of an optimal measurement outcome. The PS protocol can therefore reduce the time needed at the expense of repeated runs of the protocol to obtain the desired value of $c$ (or close to it) upon measurement. We will discuss this trade-off and variable range of $c$ in more detail.

\begin{figure}[btph]
\hspace*{-0.15cm}
  \centering
  \begin{tabular}{ccc}
\includegraphics[width=88.85mm,scale=0.99]{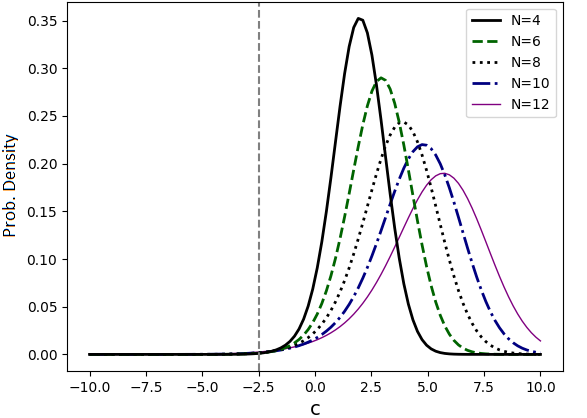}
  \end{tabular}
  \caption{The post-selection probability density functions (PDFs) for varied system sizes of $N \in \{4, 6, 8, 10, 12\}$ and $\chi t=0.15$. The dashed vertical at $c=-2.5$ denotes the identified optimal measurement, even though the distributions peak at increasing values of $c>0$, as a function of $N$.}
  \label{PDFB}
\end{figure}
The post-selection PDF for measurements $c\in[-10, 10]$ and system size $N \in \{4, 6, 8, 10, 12 \}$ is presented in Fig.~{\ref{PDFB}}. The observed trend indicates that as $N$ increases, the measurement outcome associated with the most probable result also increases. Simultaneously, the maximum probability associated with $N$ tends to decrease, while the width of the parabolic function about the maximum value increases.
\section{The Symmetric Dicke subspace}
The ideal decoherence-free PS state protocol is well-defined in the symmetric Dicke subspace (see Ref.~\cite{alexander2020generating}), spanned by the Dicke basis states
\begin{align}
\begin{split}
    \hspace{0.2cm}
    &\bigg|\frac{N}{2}, \frac{N}{2}-m\bigg\rangle \\&:= \frac{\sqrt{m!(N-m)!}}{\sqrt{N!}}\sum_{j_1<\cdot\cdot\cdot<j_m}|\uparrow \cdot\cdot\cdot\downarrow_{j_1}\cdot\cdot\cdot\downarrow_{j_m}\cdot\cdot\cdot\uparrow \rangle.
    \label{A2}
\end{split}
\end{align}
The dimension of the Dicke subspace scales linearly with system size $\sim N+1$. This favorable scaling allows access to higher $N$ as compared to modelling in the total Hilbert space, but it becomes necessary to consider the total Hilbert space when including certain decoherence channels.

In the next section, we present a quantum trajectories approach~\cite{dalibard1992wave,carlo2004simulating}, which is a numerical method for reducing the computational complexity by essentially modelling $\mathcal{N}$ independent state vector realizations, as opposed to the corresponding density operator evolution. The set of $\mathcal{N}$ pure states are then used to compute the expectation value of chosen observables. We exploit the reduction in computational complexity by evaluating, with the inclusion of decoherence channels, quantum metrological observables of interest, such as the overlap fidelity with the GHZ state, as well as the QFI. 
The convexity of the QFI as described by (\ref{conv}) yields the inequality
\begin{align}
\mathcal{Q}(\rho, \hat{H}) \leq \frac{1}{\mathcal{N}}\sum_{i=1}^{\mathcal{N}}\mathcal{Q}(\rho_i, \hat{H})=:\overline{\mathcal{Q}}(\rho, \hat{H}),
\end{align}
for some mixed state $\rho = \frac{1}{\mathcal{N}}\sum_{i=1}^{\mathcal{N}}\rho_i$, where $\mathcal{N}$ represents the number of independent trajectories. Consequently, the average QFI denoted by $\overline{\mathcal{Q}}(\rho, \hat{H})$ serves as an upper bound of the QFI. As a consequence, Eq.~(\ref{CR}) provides a lower bound in terms of the estimator variance.

Considering the computational server resources at our disposal, we have determined that a system size of $N=12$ represents an approximate upper bound for accurate modeling within a reasonable time-frame while achieving a sufficient trajectory count for the utilization of the quantum trajectory method~\cite{blogpost, johansson2012qutip, johansson2012qutip2}. As we look to the future, to simulate systems beyond $N=12$, we will explore computational software packages designed specifically for the efficient simulation of large-scale quantum systems, allowing us to potentially reach system sizes up to $N=32$~\cite{suzuki2021qulacs, zhang2023tensorcircuit, efthymiou2022quantum, CUNVid}.       

\section{The Quantum Trajectory Method}

For the purpose of numerically simulating the non-unitary dynamics of the PS state protocol with decoherence, we assume Markovian evolution described by the Lindblad form of the master equation~\cite{nielsen2002quantum, carlo2004simulating, breuer2002theory, manzano2020short}, \begin{equation}
    \frac{d \rho}{d t} = -\frac{i}{\hbar}[\hat{H}_{s}, \rho]-\frac{1}{2}\sum_{j}\big\{\hat{L}_{j}^{\dagger}\hat{L}_{j}, \rho \big\} + \sum_{j}\hat{L}_{j} \rho \hat{L}^{\dagger}_{j},
    \label{Lind}
\end{equation}
where $\hat{H}_{s}$ denotes the system Hamiltonian, and $\{ \hat{L}_{j}\}_{j \in \mathcal{M}}$ a set of Lindblad jump operators, with $\big\{\cdot, \cdot \big\}$ denoting the anti-commutator operator. The relevant decoherence channels that characterize the set of $\mathcal{M}$ jump operators are determined by the experimental setting.
The first term of Eq.~(\ref{Lind}) describes the standard unitary time-evolution of a state $\rho$, while the remaining terms generate the dissipative dynamics characterized by $\{\hat{L}_{j}\}_{j \in \mathcal{M}}$.   

Let us now consider the quantum trajectory method~\cite{carlo2004simulating}, used to simulate the preparation of the PS state with decoherence.
By defining a non-Hermitian effective Hamiltonian  
\begin{equation}
    \hat{H}_{eff}:=\hat{H}_{s}-\frac{i \hbar}{2}\sum_{j}\hat{L}^{\dagger}_{j}\hat{L}_{j}, 
    \label{Heff}
\end{equation}
we can recast the first two terms of Eq.~(\ref{Lind}) as 
\begin{equation}
    -\frac{i}{\hbar}\big[\hat{H}_{s}, \rho \big]-\frac{1}{2}\sum_{j}\big\{ \hat{L}^{\dagger}_{j}\hat{L}_{j}, \rho \big\}=-\frac{i}{\hbar}\big[ \hat{H}_{eff}\rho-\rho\hat{H}_{eff}^{\dagger}\big].
\end{equation}
An effective non-Hermitian Hamiltonian is an operator that describes the dynamics of a quantum system in the presence of decoherence~\cite{carlo2004simulating, breuer2002theory}. 
To first-order of the infinitesimal time step $dt$, the solution to Eq.~(\ref{Lind}) is given by the Kraus operator (or operator sum)~\cite{breuer2002theory, kraus1983states} formalism 
\begin{equation}
    \rho(t+dt)= \sum_{j}\hat{M}_{j}(dt)\rho(t){\hat{M}_j}^{\dagger}(dt), 
    \label{Kr}
\end{equation}
where the Kraus operators are defined as
\begin{equation}
\hat{M}_{0}:= \hat{\mathds{1}}-i\frac{dt}{\hbar}\hat{H}_{eff},\label{M0}
\end{equation}
for $j=0$, and 
\begin{equation}
\hat{M}_j := \sqrt{dt}\hat{L}_j,
\label{M1}
\end{equation}
for $j > 0$.

When including decoherence, the typical formalism used to describe the evolution of mixed quantum states is given in terms of density operators~\cite{nielsen2002quantum}. 
For large systems, simulating the $2^N \times 2^N$ density operator evolution given by Eq.~(\ref{Kr}) can become computationally expensive. This motivates the use of the quantum trajectory method, which allows us to instead independently simulate $\mathcal{N}$ pure states of dimension $2^N$.  

The numerical methods are applicable to various physical systems, but in this study our focus is on physical realization of the PS state protocol with a Penning ion trap~\cite{biercuk2009high,britton2012engineered,uys2012toward, foss2013nonequilibrium, bohnet2016quantum}.
 To this end, because the squeezing step is known to be the most susceptible to the disruptive effects of decoherence, we assume negligible decoherence during the collective rotation steps. During the spin squeezing step, the system Hamiltonian is  
\begin{equation}
    \hat{H}_s:=\hbar\chi \hat{J}_z^{2},
    \label{Hsys}
\end{equation}
where $\chi \geq 0$ regulates the squeezing magnitude (see Sec.~IIIB of Ref.~\cite{pezze2018quantum}). The total evolution time $t$ is partitioned into $N_t=t/dt$ steps. During each time step, the probability that some composite state $|\psi\rangle$ undergoes a jump (or decays) resulting from the $j$-th jump operator is given by
\begin{equation}
    dp_{j} = \langle \psi| \hat{M}^{\dagger}_{j}\hat{M}_{j}|\psi \rangle,
\end{equation}
where $j \in \mathcal{M}$. At most one jump per $dt$ time step is permitted, the validity of which depends on the form of the jump operators and their associated decay rates, in addition to how many channels of the form given in Eq.~(\ref{Lind}) are present. These factors, which influence the accuracy of the approximation of Eq.~(\ref{Lind}), also determine the required duration of the time interval, denoted as $dt$, needed to obtain a sufficiently accurate approximation. We will come back to this point later when discussing the simulations. 

Had the state $|\psi\rangle$ undergone a $j$-th jump during $dt$, the resultant state reads as  
\begin{equation}
\frac{\hat{M}_{j}|\psi\rangle}{\big\Vert \hat{M}_{j}|\psi\rangle \big\Vert},
    \label{Mj}
\end{equation}
where $\Vert \cdot \Vert$ denotes the norm. If no jump occurs, the resultant state is 
\begin{equation}
    \frac{\hat{M}_{0}|\psi\rangle}{\big\Vert \hat{M}_{0}|\psi\rangle \big\Vert}.
    \label{M00}
\end{equation}

For each $dt$ time step, the computational scheme (see Ref.~\cite{carlo2004simulating}) used to employ the quantum trajectory method proceeds by first generating a random number $\epsilon~\in~\mathcal{U}_{[0,1]}$, where $\mathcal{U}_{[0,1]}$ denotes the uniform probability distribution chosen from the interval $[0,1] \subset \mathbb{R}$. If $\epsilon \leq \sum_{j}dp_{j}$, a single $\hat{L}_j$ jump occurs, with $j>0$. The $j\in \mathcal{M}$ is chosen such that $dp_{j-1} \leq \epsilon \leq dp_{j-1}+dp_{j}$ holds. For $j=1$, we set $dp_{0} = 0$. The resultant state, given that the $j$-th jump occurred, is given by Eq.~(\ref{Mj}). Otherwise, if $\epsilon > \sum_{j}dp_{j}$, the state undergoes evolution by the non-Hermitian Hamiltonian given by Eq.~(\ref{M0}), leading to the state given by Eq.~(\ref{M00}).
This process is independently repeated for $N_t=t/dt$ iterations, for each of the $\mathcal{N}$ trajectories, yielding a set of $\mathcal{N}$ final pure states. In this way the quantum trajectory method realizes (or unravels) the master equation in Eq.~(\ref{Lind}) by stochastically evaluating the corresponding state vector $|\phi_{i} \rangle$ over a specified number of independent trajectories $i \in \{1,...,\mathcal{N}\}$, yielding $\mathcal{N}$ pure states.

As $\mathcal{N} \to \infty$, and assuming sufficiently small $dt$, averaging over the $\mathcal{N}$ trajectories recovers the classical probability coefficients of the corresponding density operator sum evolution given by Eq.~(\ref{Kr}).
For the PS state protocol, to recover the final PS state density operator from the corresponding set of pure state realizations, we are required to weight each pure state by its respective post-selection conditional probability 
\begin{align}
\text{P}(i|c) = \text{P}(c|i)\text{P}_i/\text{P}(c),
\end{align}
 with $\text{P}_i=1/\mathcal{N}$,  $\text{P}(c|i) = \langle \phi_i | \hat{A}_c^{\dagger}\hat{A}_c|\phi_i\rangle$ and $\text{P}(c) = \sum_{i}\text{P}(c|i)\text{P}_i$, where $i$ is the trajectory count and $c$ the measurement outcome (see Appendix B for further details).

To achieve an acceptable level of statistical precision, the required number of trajectories $\mathcal{N}$ is taken as the approximate number for which the sample standard deviation (SD) of the set of overlap fidelities $\mathcal{F}$, stabilizes around certain fixed values. By choosing a suitable trajectory count $\mathcal{N}$, we are afforded some control over the total computation time. This is a useful computational feature when simulating large systems. With each trajectory being independent, the quantum trajectory method lends itself well to the implementation of parallel computing techniques and the utilization of specialized computing infrastructure. This reduction in computation time enables access to larger system sizes.   
  
To approximate the expectation value of some Hermitian operator $\hat{A}$, we consider the mean of the expectation values of the $\mathcal{N}$ pure states. The expectation value of $\hat{A}$ follows by taking the limit as $\mathcal{N}\to \infty$,
\begin{equation}
    \langle \hat{A} \rangle = \text{Tr}(\hat{A}\rho) = \lim\limits_{\mathcal{N} \to \infty}\frac{1}{\mathcal{N}}\sum_{i=1}^{\mathcal{N}}\langle \tilde{\phi}_i|\hat{A}|\tilde{\phi}_i\rangle,
    \label{Exp}
\end{equation}
where $|\tilde{\phi}_i \rangle$, defined by $|\phi_i\rangle$ weighted by $\text{P}(i|c)$, i.e., $|\tilde{\phi}_i\rangle=\sqrt{\text{P}(i|c)}|\phi_i\rangle$ denotes the state vector realized by the $i$-th trajectory after measurement outcome $c$ is obtained. 

\section{Numerical Analysis}

Quantum decoherence, in general, tends to degrade quantum correlations. In Ref.~\cite{foss2013nonequilibrium}, a study is undertaken to investigate the effects of decoherence on an $N$-body quantum system. The system is subjected to an assumed long-range Ising Hamiltonian, which we will use for our spin squeezing step. The Hamiltonian is defined as
\begin{equation}
\hat{H}:=\frac{1}{N}\sum_{i<j}\hbar J_{ij}\hat{\sigma}^{z}_{i}\hat{\sigma}^{z}_{j},
\label{foss}
\end{equation}
where
\begin{equation}
J_{ij}:=J|\vec{r}_i-\vec{r}_j|^{-\zeta}
\label{Jij}
\end{equation}
parameterizes the spin-spin interaction with chosen scalar parameters $J$ and $\zeta$. The $i$-th qubit position vector, in lattice units, is denoted by $\vec{r}_i$. Trapped-ion systems with $0 < \zeta < 3$ have been shown to be a feasible means of realizing the Hamiltonian specified by Eq.~(\ref{foss}) (see Refs.~\cite{porras2004effective, jurcevic2014quasiparticle, bohnet2016quantum, plodzien2020producing}). We set $J_{ij} = J$ for $\zeta=0$, which is implemented by an optical dipole force tuned to be nearly resonant with the center-of-mass mode of a trapped-ion crystal~\cite{bohnet2016quantum}. For the purpose of simulating these types of experimental setups, we consider the dephasing, amplitude damping and spontaneous excitation channels, which we will define by their respective state-bath mappings in the next section. The Hamiltonian presented in Eq.~(\ref{foss}) constitutes the essential interaction for the spin squeezing step within the PS state protocol (refer to Eqs.~(\ref{Hsys}) and (\ref{Sq})). We will provide a clearer explanation of the relationship between Eqs.~(\ref{foss}) and (\ref{Hsys}) after introducing the decoherence channels. 

It is important to note that Eq.~(\ref{foss}) is a simplification of a transverse-field Ising spin model, which usually has an additional term $\hat{H}_{B}=\sum_{i=1}^{N}B_{\mu}\hat{\sigma}_{i}^{\mu}$ (see Refs.~\cite{bohnet2016quantum, elliott1970ising}), where $B_{\mu}$ is a transverse magnetic field in the $\mu$-direction. Usually $B_x$ and $B_y$ are used for implementing rotations and set to zero during the spin squeezing step of the protocol which we are concerned with. The $B_z$ term represents a local contribution originating from the magnetic field of the trap, leading to the energy splitting of the ion's spin states. In our simulation, we utilize $\hat{H}$ in Eq.~(\ref{foss}) within the rotating frame, excluding the $B_{z}$ term. In Appendix C we show that the measurements $\hat{A}_c$ are equivalent in the rotating frame and original/lab frame. The decoherence channels we consider also do not change the form of Eq.~(\ref{Lind}) in this rotating frame~\cite{preskill2015lecture}. 

In an experiment, collective rotations about the x or y-axis can be performed in the rotating frame of the spins by setting the phase of the applied resonant microwave field~\cite{wolfowicz2016pulse,
bardin2021microwaves}. Thus, there is no need to adjust these rotations for the lab frame. 

\subsection{Dephasing}

The Kraus operators of Eqs.~(\ref{M0}) and (\ref{M1}) are characterized by the chosen decoherence channels.
The dephasing channel, which is equivalent to the phase flip channel of a single qubit coupled to a bath, is defined by the state-bath mappings 
\begin{multline*}
~~|0\rangle_{S} \otimes |0\rangle_{B} \mapsto \sqrt{1-p}|0\rangle_{S}\otimes|0\rangle_{B}+\sqrt{p}|0\rangle_{S}\otimes|1\rangle_{B}
\end{multline*}
and
\begin{multline}
|1\rangle_{S} \otimes |0\rangle_{B} \mapsto \sqrt{1-p}|1\rangle_{S}\otimes|0\rangle_{B}-\sqrt{p}|1\rangle_{S}\otimes|1\rangle_{B},
\label{sbdep}
\end{multline}
where we assume that the bath is initially in the state $|0\rangle_{B}$, with probability $p = \Gamma_{\text{dep}}dt$ that a phase flip occurs during a $dt$ time step, where $\Gamma_{\text{dep}}$ is the dephasing rate. Tracing out the bath yields the corresponding dephasing Kraus operators 
\begin{equation}
\hat{M}_{0} = 
\begin{pmatrix}
\sqrt{1-p} & 0 \\
0 & \sqrt{1-p} 
\end{pmatrix}~~\text{and}~~\hat{M}_{1}=\begin{pmatrix}
\sqrt{p} & 0 \\
0 & -\sqrt{p}
\end{pmatrix}.
\label{Kdep}
\end{equation}
The dephasing state-bath mappings given by Eq.~(\ref{sbdep}) is an effective model, 
where $|0\rangle_{B}$ and $|1\rangle_{B}$ denote 
orthogonal states that result from an elastic scattering process with no energy transfer. 
 Ref.~\cite{preskill2015lecture} provides the full model, with a more rigorous description.

This Kraus operator method for a single qubit approximates the master equation in Eq.~(\ref{Lind}) with a single jump operator $\hat{L}_j$ obtained by substituting $\hat{M}_1$ in Eq.~(\ref{M1}). The approximation is valid assuming $dt$ is much less than the time scales relevant for the evolution of the system $\sim 1/\Gamma_{\text{dep}}$. In other words, if $\Gamma_{\text{dep}}dt = p \ll 1$, one can then apply the channel in Eq.~(\ref{Kr}) iteratively to recover the dynamics over time $t$. The same approach is also utilized for the amplitude damping and spontaneous excitation channels, with their respective definitions presented next. This Kraus operator method is used for low $N$. For high $N$ we employ the quantum trajectory method.

\subsection{Amplitude damping and spontaneous excitation}

The amplitude damping (spontaneous deexcitation) channel of a single qubit coupled to a bath is defined by the following state-bath mappings
\begin{gather*}
\centering
|0\rangle_{S} \otimes |0\rangle_{B} \mapsto |0\rangle_{S}\otimes|0\rangle_{B}
\end{gather*}
and
\begin{multline}
|1\rangle_{S} \otimes |0\rangle_{B} \mapsto \sqrt{1-p}|1\rangle_{S}\otimes|0\rangle_{B}+\sqrt{p}|0\rangle_{S}\otimes|1\rangle_{B},
\label{sbad}
\end{multline}
where again we assume that the bath is initially in the vacuum state $|0\rangle_{B}$ and $p = \Gamma_{\text{ad}}dt$, with $\Gamma_{\text{ad}}$ the amplitude damping rate. Tracing out the bath yields the corresponding amplitude damping Kraus operators 
\begin{equation}
\hat{M}_{0} = 
\begin{pmatrix}
1 & 0 \\
0 & \sqrt{1-p} 
\end{pmatrix}~~\text{and}~~\hat{M}_{1}=\begin{pmatrix}
0 & \sqrt{p} \\
0 & 0 
\end{pmatrix}.
\label{depm}
\end{equation}

The spontaneous excitation channel~\cite{foss2013nonequilibrium} of a single qubit coupled to a bath is defined by the state-bath mappings
\begin{gather*}
\centering
|1\rangle_{S} \otimes |1\rangle_{B} \mapsto |1\rangle_{S}\otimes|1\rangle_{B}
\end{gather*}
and
\begin{multline}
|0\rangle_{S} \otimes |1\rangle_{B} \mapsto \sqrt{1-p}|0\rangle_{S}\otimes|1\rangle_{B}+\sqrt{p}|1\rangle_{S}\otimes|0\rangle_{B},
\label{sbse}
\end{multline}
where $p = \Gamma_{\text{se}}dt$, with $\Gamma_{\text{se}}$ denoting the spontaneous excitation rate.
Tracing out the bath yields the corresponding spontaneous excitation Kraus operators 
\begin{equation}
\hat{M}_{0} = 
\begin{pmatrix}
\sqrt{1-p} & 0 \\
0 & 1 
\end{pmatrix}~~\text{and}~~\hat{M}_{1}=\begin{pmatrix}
0 & 0 \\
\sqrt{p} & 0 
\end{pmatrix}
\label{sek}.
\end{equation}

For each of the decoherence channels introduced, when the Kraus operator method is used, $\hat{M}_0$ is modified in the presence of spin squeezing in the protocol. This is taken into account by redefining it using Eq.~(\ref{M0}). More details regarding the multi-qubit case are given at the end of this section. Typical experimental parameters for the NIST Penning ion trap work, necessary for numerically simulating the unitary spin squeezing steps of the PS state and MSS protocols with decoherence, are provided in Ref.~\cite{foss2013nonequilibrium}. 
Note that $\chi dt \ll 1$ should be used if $\chi > \Gamma$, where $\chi$ is the strength of some operation, for Eq.~(\ref{Lind}) to be well-approximated e.g., the spin squeezing strength in Eq.~(\ref{Hsys}). This also ensures $\Gamma dt \ll 1$ (see Ref.~\cite{preskill2015lecture}).

For the dephasing, amplitude damping, and spontaneous excitation channels, we employ decay rates consistent with experimental work conducted at NIST, where these decay rates, along with the Hamiltonian parameter $J$ mentioned in Eq.~(\ref{Jij}) (see Appendix D and Refs.~\cite{foss2013nonequilibrium, britton2012engineered}) are related by 
 \begin{equation}
    4\Gamma_{\text{dep}}+ \Gamma_{\text{ad}}+\Gamma_{\text{se}}=J/10,
 \label{Gam1}
 \end{equation}
with $\Gamma_{\text{ad}}= \Gamma_{\text{se}}$ and $\Gamma_{\text{dep}} = 2\Gamma_{\text{ad}}$, which leads to 
\begin{equation}
\begin{split}
    \{\Gamma_{\text{dep}},\Gamma_{\text{ad}},\Gamma_{\text{se}}\}
     = \{J/50,J/100,J/100\}.
    \label{Gam2}
\end{split}
\end{equation}
As such, it is clear that dephasing is the largest type of decoherence. The specifications, including laser detuning and polarizations, that are specific to the NIST Penning ion trap setup are addressed by Eqs.~(\ref{Gam1}) and (\ref{Gam2}), as detailed in Refs.~\cite{foss2013nonequilibrium, britton2012engineered}.
 
The commutativity between the dephasing operator, as given by Eq.~(\ref{depm}), and the spin squeezing operator, as given by Eq.~(\ref{Sq}), respectively, is particularly noteworthy. Consequently, the presence of amplitude damping and spontaneous excitation channels requires the adoption of the Kraus operator sum and quantum trajectory methods to effectively account for these effects. 

In Ref. \cite{bohnet2016quantum}, no evidence was found to suggest a significant role of motional decoherence. The primary source of decoherence was identified as off-resonant light scattering. While our study does not include an analysis of motional decoherence, it is anticipated to only be relevant during the application of single-axis twisting. Longer durations of single-axis twisting are expected to exhibit greater sensitivity to motional decoherence compared to shorter durations. Consequently, incorporating an analysis of motional decoherence is expected to strengthen the argument that the PS state protocol offers resilience against environmental effects as the number of qubits increases.
 
 By reconciling the spin squeezing Hamiltonian given in Eq.~(\ref{Hsys}), with that of Eq.~(\ref{foss}), we find the following relation between the spin squeezing and Hamiltonian parameters
 \begin{equation}
    \chi = 2 J/N. 
 \label{chi}
 \end{equation}
The parameter $J$ only depends on laser parameters~\cite{bohnet2016quantum}. For fixed laser parameters, Eq.~(\ref{chi}) implies that an increase in $N$ results in a decrease in the squeezing parameter $\chi$. Hence, for a fixed $\chi t$ (e.g., $\pi/2$ for MSS) we require an increase in the total squeezing time~$t$, as $N$ increases.

For the dephasing channel, defined by the state-bath mappings of Eq.~(\ref{sbdep}), the decay probability per-qubit of an $N$-qubit system during a $dt$ time step, in terms of the squeezing parameter $\chi$ follows by considering Eq.~(\ref{chi}) in Eq.~(\ref{Gam2}), leading to 
\begin{equation}
    p_{1}(x) := x \Gamma_{\text{dep}} dt = \frac{x N \chi}{100}dt,
    \label{dec}
\end{equation}
where we introduce the decay parameter $x\geq0$ to control the magnitude of the decay probability. The values of $x$ used in this study, specifically $x=1$, align with the experimentally motivated values mentioned in Refs.~\cite{foss2013nonequilibrium, britton2012engineered}. Similarly, the decay probability per-qubit of the amplitude damping and spontaneous excitation channels, defined respectively by Eqs.~(\ref{sbad}) and (\ref{sbse}), reads as
\begin{equation}
    p_{2}(x) :=  \frac{x N \chi}{200}dt,
    \label{dec3}
\end{equation}
and thus $p_{2}(x)=\frac{1}{2}p_{1}(x)$.  Hereafter, we assume that a chosen $x$ uniformly applies to all channels. We also assume the experimentally relevant value of $J=3300~\text{s}^{-1}$. This value was found by calibrating $J$ through measurements of the mean-field spin precession~\cite{bohnet2016quantum}.

For simulating the PS state and MSS generation with the inclusion of relevant decoherence channels during the spin squeezing step of each protocol, we utilized both the Kraus operator and quantum trajectory methods. In cases where the computational feasibility of the Kraus operator evolution of Eq.~(\ref{Kr}) was constrained, we exclusively employed the quantum trajectory method, particularly for $N \geq 12$. For the Kraus method, as we have $N$ qubits, we assume each qubit has its own independent bath of the form given by Eqs.~(\ref{Kdep}), (\ref{depm}) and (\ref{sek}), and that at most one qubit should experience a jump per $dt$ time step. In general, the product of $N$ channels of the form given in Eq.~(\ref{Kr}) can give rise to terms with multiple jumps per time step. However, for the channels described by Eqs.~(\ref{Kdep}), (\ref{depm}), and (\ref{sek}), there are a total of $N$ terms corresponding to individual jumps. Each of these terms represents the probability, denoted as $\Gamma dt$, that a single qubit will undergo the jump. Therefore the total probability for a single jump to occur for a given channel is $N \Gamma dt$, and the probability for two jumps is $\text{C}(N,2)(\Gamma dt)^2$, where $\text{C}(N,2) = \frac{N!}{2!(N-2)!}$. Thus the relative probability for two jumps to occur compared to one jump is $(N-1)\Gamma dt/2$. We restrict our consideration to $N \leq 12$, and choose $\Gamma dt=p$ such that the probability of more than one jump per time step $dt$ in a single channel is negligible. This approach is equivalent to the independent channel model given in Ref.~\cite{carlo2004simulating}, which approximates the master equation in Eq.~(\ref{Lind}) when written as a summation of jump operators acting independently on each qubit for each channel type.

Practically, the above is achieved by using $\hat{M}_{j}$ for $j>0$ in Eq.~(\ref{Kr}), formed by taking all $N$ permutations of $\hat{M}_{1}^{(1)} \otimes \hat{\mathds{1}}^{(2)}\otimes\cdot\cdot\cdot\otimes  \hat{\mathds{1}}^{(N)}$ from each damping channel, and $\hat{M}_0$ from Eqs.~(\ref{M0}) and~(\ref{Heff}), i.e., 
\begin{align}
\rho(t+dt) =&  \hat{M}_{0}\rho(t)\hat{M}_{0}^{\dagger} \notag \\
&+ \hspace{-3ex} \sum_{\text{perm. of dep.}}\hat{M}_{1} \otimes \hat{\mathds{1}} \cdot\cdot\cdot \hat{\mathds{1}}\rho(t)(\hat{M}_{1} \otimes \hat{\mathds{1}} \cdot\cdot\cdot \hat{\mathds{1}})^{\dagger} \notag \\
&+ \hspace{-3ex} \sum_{\text{perm. of ad.}}\hat{M}_{1} \otimes \hat{\mathds{1}} \cdot\cdot\cdot \hat{\mathds{1}}\rho(t)(\hat{M}_{1} \otimes \hat{\mathds{1}} \cdot\cdot\cdot \hat{\mathds{1}})^{\dagger} \notag \\
&+ \hspace{-3ex} \sum_{\text{perm. of se.}}\hat{M}_{1} \otimes \hat{\mathds{1}} \cdot\cdot\cdot \hat{\mathds{1}}\rho(t)(\hat{M}_{1} \otimes \hat{\mathds{1}} \cdot\cdot\cdot \hat{\mathds{1}})^{\dagger}.
\end{align}
For the quantum trajectory method, we use the same $\hat{M}_{j}$ as in the Kraus method, but follow the steps described in Eqs.~(\ref{Hsys})--(\ref{M00}), again allowing only one jump per time step which gives an unraveling of the Kraus and master equation approaches. This approach yields a set of pure states with vectors of size $2^N$ (as opposed to matrices of size $2^N \times 2^N$ in the Kraus method), and as such allows access to the dynamics of larger systems.
\begin{figure}
\hspace*{-0.28cm}
  \centering
  \begin{tabular}{ccc}
\includegraphics[width=84mm,scale=0.99]{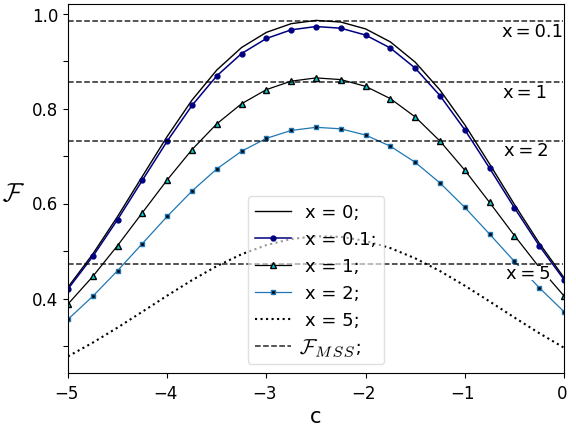}
  \end{tabular}
  \caption{Evaluating $\mathcal{F}$ and $\mathcal{F}_{\text{MSS}}$ (the dashed horizontals) with the Kraus operator formalism given by Eq.~(\ref{Kr}), where $x\in \{0, 0.1, 1, 2, 5\}$ for $N=4$ (see Eqs.~(\ref{dec}) and (\ref{dec3})).}
  \label{4q}
\end{figure}
\begin{figure}
\hspace*{-0.28cm}
  \centering
  \begin{tabular}{ccc}
\includegraphics[width=84.5mm,scale=0.99]{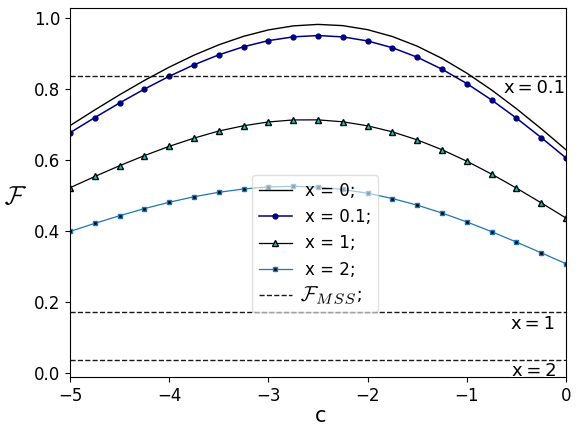}
  \end{tabular}
  \caption{Evaluating $\mathcal{F}$ and $\mathcal{F}_{\text{MSS}}$ (the dashed horizontals) with the Kraus operator formalism given by Eq.~(\ref{Kr}), where $x\in \{0, 0.1, 1, 2\}$ for $N=10$ (see Eqs.~(\ref{dec}) and (\ref{dec3})).}
  \label{10q}
\end{figure}
\subsection{Computational Results}
An analysis of a PS state with $N = 4$ and $N = 10$, generated using $\chi t = 0.15$, and $x \in \{0, 0.1, 1, 2, 5\}$ as defined by Eqs.~(\ref{dec}) and (\ref{dec3}), is presented in Fig.~\ref{4q} and Fig.~\ref{10q}, respectively. Moreover, we compare the overlap fidelity of the PS state and GHZ state, denoted by $\mathcal{F}$, with the overlap fidelity of the MSS and GHZ state, hereafter denoted by $\mathcal{F}_{MSS}$. For the PS state, we observe a symmetry of the fidelity about the optimal identified measurement of $c \approx -2.5$. There is a clear downward monotonic trend as the decay rate $x$ increases from $x=0$ to $x=5$ for the PS and MSS generating protocols. For increasing values of $x \geq 1$, we observe an increasingly superior fidelity, $\mathcal{F} > \mathcal{F}_{MSS}$, demonstrated by the PS state compared to the MSS. A notable enhancement in the relative performance of the PS state protocol is observed when scaling the system size from $N=4$ to $N=10$, compared to the MSS protocol.

The measurement step of the PS state protocol distinguishes it from the deterministic MSS generation by introducing non-unitary evolution, as such, we must consider the associated post-selection probabilities. For $N \in \{4,6,8,10,12\}$, Table~\ref{TAB} in Appendix E presents a post-selection analysis of the PS state generation with $x \in \{0, 1\}$, as well as the corresponding $\mathcal{F}_{MSS}$ for comparison. We employ the Kraus operator method, incorporating decoherence channels, to investigate the PS state protocol for system sizes up to $N = 10$. Additionally, we examine the $N = 12$ case without decoherence ($x = 0$). To account for decoherence in the $N = 12$ case, we utilized the quantum trajectory method. In Table~\ref{TAB}, we analyze various measurement intervals of the form $[-2.5, \cdot]$. These asymmetric intervals around the optimal measurement outcome, $c = -2.5$, are chosen due to the known asymmetry in the post-selection probability density functions (PDFs), as depicted in Fig.~\ref{PDF} and Fig.~\ref{PDFB}, for spin squeezing of $\chi t = 0.15$. Table~\ref{TAB} suggests that for decay parameters $x \in \{0, 1\}$, the success rate of post-selecting measurements $c \in [-2.5,-1.5]$ remain relatively constant at about $0.004$ for $N \in \{4,6,8,10,12\}$. For each considered $N$, there is a slight increase observed in obtaining the desired post-selected outcome from $[-2.5, 1.5]$ when including decoherence with $x = 1$, relative to no decoherence ($x = 0$). The minimum and maximum fidelity values obtained for the specified measurement intervals tend to decrease with the inclusion of decoherence channels for all considered $N$.  

By post-selecting on a sufficiently small interval, it is clear that $\mathcal{F}>\mathcal{F}_{MSS}$ can be obtained for all $N$ when including relevant decoherence channels with $x=1$, where the superior fidelity becomes more pronounced with increasing system size $N$ (see Table~\ref{TAB}). This further promotes PS states as robust entangled resources that are practically feasible, as their efficacy improves when scaling to larger systems. 

In Table~\ref{TAB}, consider the results for $N=10$, $x=1$, and $c \in [-2.5, 1.5]$. Notably, the evolution time for the PS state is $0.15/(\pi/2) \approx 1/10$ of that required by the MSS, while the success rate for post-selecting from $[-2.5, 1.5]$ is also approximately $1/10$. Assuming that the time interval required for evolution by single-axis twisting is considerably longer than the other steps in the protocol\footnote{For the NIST experimental set-up in Ref.~\cite{bohnet2016quantum} the duration of the single-axis twisting step for $\chi t = 0.15$ and $N = 10$ is about 5\% of the duration of a single experiment. However, the duration of other steps can be reduced in principle, making the assumption valid. Furthermore, in general the single-axis twisting time grows linearly with $N$, while other steps, such as motional cooling and spin-state detection have times independent of $N$.}, the rate of obtaining a GHZ-like state with the PS protocol is approximately equivalent to the unitary evolution of the MSS. Moreover, the fidelity of the PS state post-selected from $[-2.5, 1.5]$ varies between a minimum of $0.20$, and maximum of $0.71$, exceeding the fidelity of $0.17$ for the MSS. Due to the reduced one-axis twisting requirements, the PS state's superior robustness allows for a more efficient generation of higher fidelity GHZ-like states, even when obtaining less favorable post-selection results from this interval. The comparative advantage is anticipated to become more pronounced with increasing $N$. We will discuss this in more detail in subsequent sections.

\begin{figure}[tbp!] \hspace*{-0.575cm}
    \centering 
    \begin{tabular}{cc} 
        \includegraphics[width=0.255\textwidth]{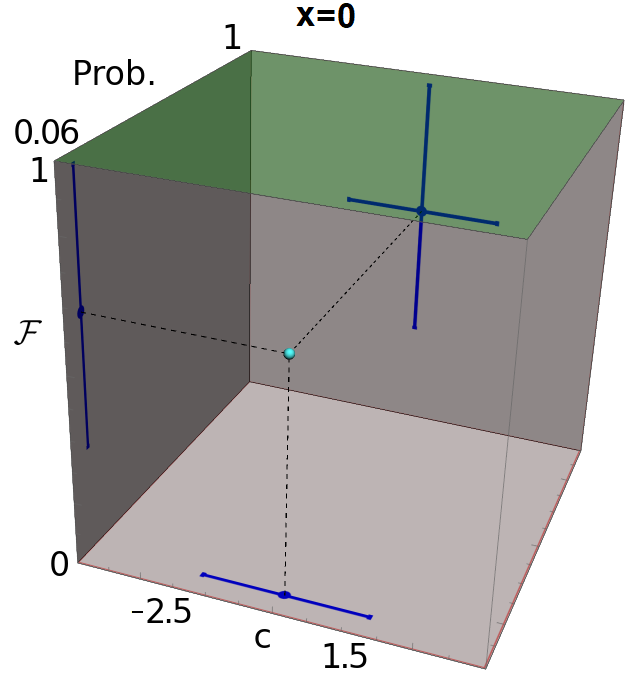} &  
        \hspace*{-0.31cm}
        \includegraphics[width=0.265\textwidth]{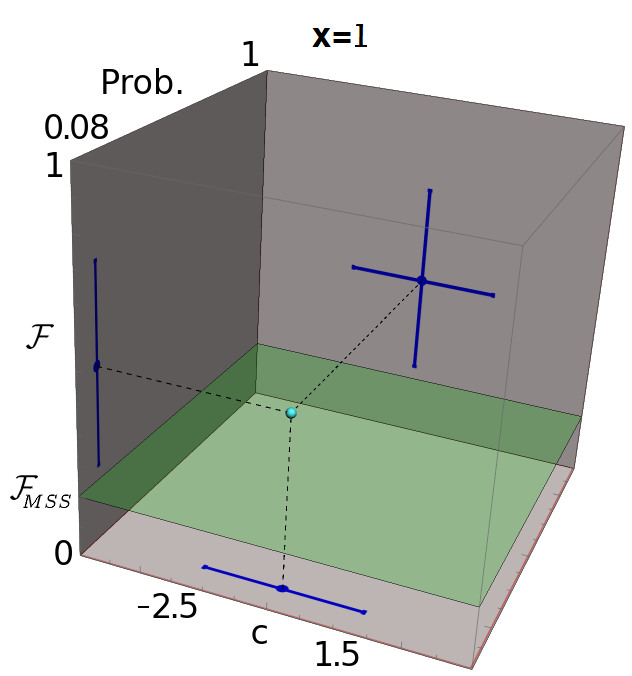}  
    \end{tabular}
  \caption{Post-selection analysis for $x=0$ (left) and $x=1$ (right) for $N=10$. The point denotes the median of the measurement interval $[-2.5, 1.5]$, the median of the resultant $\mathcal{F}$, and the post-selection probability $\text{Pr}[-2.5\leq c \leq 1.5]$. The planes denote the $\mathcal{F}_{MSS}$ values of $1$ and $0.17$. The vertical and horizontal bars represent intervals and are not error bars.}
  \label{N4Tab} 
\end{figure}
\begin{figure}
\hspace*{-0.2cm}
  \centering
  \begin{tabular}{ccc}
\includegraphics[width=90mm,scale=0.99]{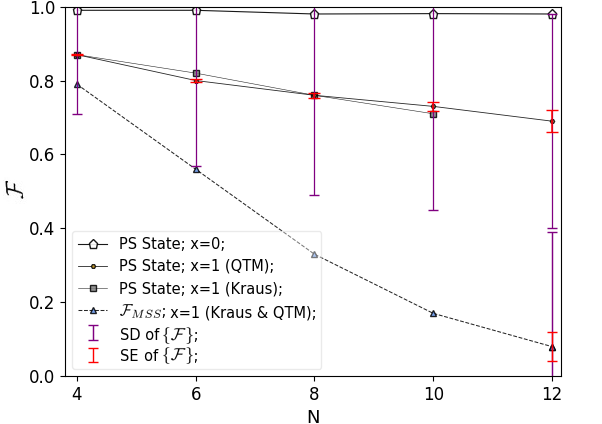}
  \end{tabular}
   \caption{The Kraus operator and quantum trajectory method, used for a $\mathcal{F}$~vs.~$\mathcal{F}_{MSS}$ study with $c=-2.5$, where $N \in \{4, 6, 8, 10, 12\}$ and $x \in \{0, 1\}$. For $N=12$, to simulate the PS and MSS generating protocols with $x=1$, we utilized the quantum trajectory method with $\mathcal{N}=200$. Thus, there are error bars (SD and SE) for the final point of the MSS plot ($N=12$). The `SD of $\{\mathcal{F}\}$' denotes fidelity spread among trajectories, $\mathcal{F}$ represents the mean fidelity for each $N$, and the `SE of 
$\{\mathcal{F}\}$' quantifies sampling variation in mean fidelity across trajectory sets.}
  \label{5q}
\end{figure}
In Fig.~\ref{N4Tab}, we depict the post-selection results of $N=10$ for measurements $c \in [-2.5, 1.5]$,
for $x = 0$ (left plot), and $x=1$ (right plot). The coordinates of the point respectively denote the median $c = -0.5$ of the chosen measurement interval $[-2.5,1.5]$ (represented by the horizontal bars), the median of the resultant fidelity set $\big\{\mathcal{F}_c:c \in [-2.5,1.5]\big\}$, where the vertical bars represent the corresponding range of fidelity values, and finally, the probability of post-selecting from $[-2.5,1.5]$, which take values of $0.06$ and $0.08$ as indicated on the respective plots. The horizontal planes represent the comparative $\mathcal{F}_{MSS}$ values of $1$ and $0.17$.  

\subsection{The Kraus operator and quantum trajectory method for $\mathcal{F}$ and $\mathcal{F}_{MSS}$}
A comparative fidelity study is shown in Fig.~\ref{5q}, where $\mathcal{F}$ and $\mathcal{F}_{MSS}$ are evaluated using the Kraus operator and quantum trajectory methods. This evaluation includes the dephasing, amplitude damping, and spontaneous excitation channels during spin squeezing. For all examined values of $N$, we observe that $\mathcal{F}>\mathcal{F}_{MSS}$ when $x=1$ and the post-selected measurement outcome is set optimally at $c=-2.5$. Notably, this difference grows with increasing~$N$. 

The quantum trajectory method enables us to obtain results, with decoherence included, for system sizes as large as $N=12$. Depending on the system size being evaluated and the associated computation time, we assume a trajectory number of $200 \leq \mathcal{N} \leq 6000$. As we obtain $\mathcal{F}$ values from the independent trajectories, we evaluate the corresponding sample standard deviation (SD) of this set, up to and including the trajectory being evaluated (the final value is represented by the error bar `SD of $\{\mathcal{F}\}$'). The relatively large final standard deviation of the set of $\mathcal{F}$ values shown in Fig.~\ref{5q} for the PS state protocol using the quantum trajectory method when there is decoherence ($x=1$) is a result of fluctuations of the obtained fidelity values, indicating a certain level of sensitivity to the decaying effects of the decoherence channels. The mean of $\mathcal{F}$ for the sample, which we simply denote by $\langle \mathcal{F} \rangle$, is the fidelity of the final state $\rho$. Similarly, we have a running evaluation of $\text{SD}/\sqrt{\mathcal{N}}$, known as the standard error (SE) and represented by the error bars `SE of $\{\mathcal{F}\}$'. The standard error is a measure of the variability in the estimate of the overlap fidelity statistic due to the natural variation that occurs when taking multiple samples from the same population (see Sec.~IIIB of Ref.~\cite{daley2014quantum}). An acceptable level of statistical precision, and thus a suitably high $\mathcal{N}$, is indicated by yielding consistent SD values with minimal deviation.

The iterative process stops when the SD reaches a steady-state criterion. The number of trajectories required for sufficient convergence depends on the system's specifics, but is typically a trajectory count that yields $\text{SE}/\langle \mathcal{F} \rangle \ll 1$ (see Ref.~\cite{daley2014quantum}). In practical applications, it is crucial to also take into account the computational resources at hand.  

\subsection{The quantum trajectory method for the QFI}
As shown in the preceding section, when accounting for decoherence, the class of PS states outperforms the MSS unitary generation scheme in fidelity, thereby establishing it as a promising resource for phase estimation in quantum metrology. This task involves estimating a unitary parameter $\theta \ll 1$ of a state that has undergone some unitary evolution as given by Eq.~(\ref{UH}).
To this end, we assume that after generating the PS state (and independently the MSS), it undergoes a unitary evolution which encodes the phase~$\theta$.
Assuming the unitary transformation is characterized by the  Hamiltonian operator $\hat{H} := \hat{J}_z$, the GHZ state is known to achieve the Heisenberg limit for $(\Delta \theta)^2$ (see Ref.~\cite{toth2014quantum}), due to the maximum possible QFI attainable for all $N$,   
    \begin{align}
        \mathcal{Q}\big(|GHZ\rangle \langle GHZ |\big)= N^2.
    \label{QFIGHZ}
    \end{align}   
    
\begin{figure}
\hspace*{-0.3cm}
  \centering
  \begin{tabular}{ccc}
\includegraphics[width=88.25mm,scale=0.99]{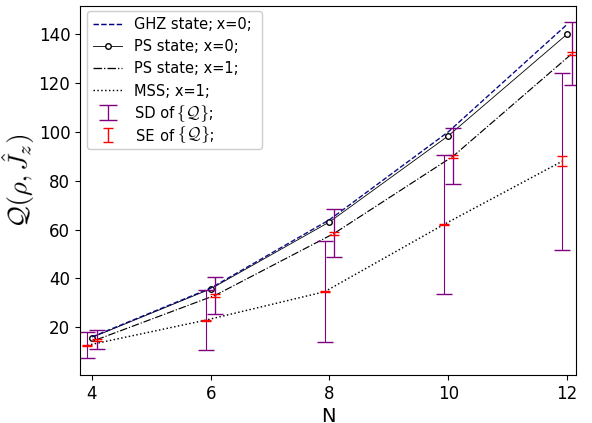}
  \end{tabular}
  \caption{A comparative QFI study of the PS and MSS generating protocols with $c=-2.5$, where $N \in \{4, 6, 8, 10, 12\}$ and $x \in \{0, 1\}$, using the quantum trajectory method with $200~\leq~\mathcal{N} ~\leq~6000$. The `SD of $\{\mathcal{Q}\}$' denotes QFI spread among trajectories, $\mathcal{Q}(\rho, \hat{J}_z)$ represents the mean QFI for each $N$, and the `SE of 
$\{\mathcal{Q}\}$' quantifies sampling variation in mean QFI across trajectory sets.}
  \label{QFIp}
\end{figure}

In this context, Fig.~\ref{QFIp} compares the QFI of the PS states, MSSs and GHZ states for $N \in \{4,6,8,10,12\}$, taking into account the effects of decoherence as described by Eqs.~(\ref{Kdep}), (\ref{depm}) and (\ref{sek}) with $x=1$. In line with the fidelity analysis, we ascertain the sample standard deviation values for the QFI set while computationally examining the trajectories.
It is clear that the class of PS states demonstrate a consistently superior QFI, and are therefore more suitable resource states for phase estimation. 

To summarize, our findings suggest that PS states exhibit greater resilience to the effects of experimentally relevant decoherence than MSSs, as demonstrated by consistently higher GHZ fidelity and QFI outcomes.

\section{Discussion}
For an $N$-qubit system with an assumed infinite range Ising interaction, characterized by $\zeta = 0$ in Eq.~(\ref{Jij}), the required time for quantum correlations to become significant scales with system size $N$ (see Ref.~\cite{foss2013nonequilibrium}) as $t \sim N^{1/3}$ for spin squeezing, $t \sim N^{1/2}$ for transverse-spin relaxation, and $t \sim N$ for the generation of MSSs. The PS state protocol works to a large extent due to transverse spin relaxation generated by one-axis twisting, which causes the state to wrap around the Bloch sphere for $\chi t \gtrsim N^{-1/2}$ (see Sec.~III of Ref.~\cite{pezze2018quantum}). With the evolution of $\chi t = 0.15$ used by the modified protocol, we find that a full wrap around the Bloch sphere is not required. While we have used $\chi t = 0.15$ in this study, the transverse spin relaxation time relative to the MSS state generation time decreases as $N$ increases. The optimal projected squeezed state evolution time is expected to scale with that of the transverse spin relaxation with $t_{PS} \lesssim N^{1/2}/2J$, where $\chi = 2J/N$. On the other hand, for MSSs the optimal time is $t=\pi/2 \chi$ (see Ref.~\cite{pezze2018quantum}), which gives $t_{MSS} = \pi N/4J$. Therefore, the ratio of the optimal evolution times for the PS state and the MSS generation, scales as $\sim N^{1/2}/N = N^{-1/2}$.

As a result, with the inclusion of relevant decoherence channels, the PS state protocol is expected to show improved efficacy, with increasing $N$, in producing GHZ-like states with higher fidelity when compared to the unitary MSS generation. The numerical results suggest that this advantage becomes more pronounced when increasing the per-qubit decay rates given by Eqs.~(\ref{dec}) and (\ref{dec3}). This is a consequence of the PS state protocol's reduced evolution time, when compared to that of the MSS. The significantly shorter evolution time produces a state with lower-order spin correlations compared to the MSS, making them less sensitive to decoherence. It is important to acknowledge that although MSSs possess a deterministic natural advantage, the more pronounced impact of decoherence offsets this advantage. In the presence of decoherence, the class of PS states exhibit superior robustness. 

 This work serves as a motivation for exploring studies involving even larger $N$.~Although such studies might seem prohibitively challenging for general forms of spin decoherence, restricting spin decoherence to dephasing, which commutes with the Ising $\hat{\sigma}^z_{i}\hat{\sigma}^z_{j}$ interaction in Eq.~(\ref{foss}), could facilitate investigations with significantly larger $N$ (see Ref.~\cite{foss2013nonequilibrium}). Also, more comprehensive studies with larger levels of decoherence can further document the apparent advantage of the  PS protocol observed in Figs.~\ref{4q} and \ref{10q}, as the level of decoherence is increased.

In summary, for the purpose of extending the PS state protocol to the total Hilbert space, improving its efficiency by reducing the required spin squeezing, and ultimately yielding PS states with more favorable GHZ overlap fidelity and QFI, we presented a modification of the protocol described in Ref.~\cite{alexander2020generating}. By including an additional collective $\hat{J}_{x}$ rotation, a reduced squeezing magnitude of $\chi t \approx 0.15$ became feasible for low $N$, as opposed to the assumed $\chi t \approx 0.40$ of Ref.~\cite{alexander2020generating}. These modifications yield PS states with superior $\mathcal{F}$ and QFI.
 By employing numerical methods we were able to reduce the required computational resources and in doing so identified optimal measurement parameters for producing high fidelity PS states for varied system sizes. Our comparative assessment of the QFI further highlights the utility of the PS state as a resource for phase estimation schemes. More generally, our findings promote PS states as viable resources for quantum information processing applications.

\section{Acknowledgements}
We acknowledge the South African Council for Scientific and Industrial Research (CSIR), and the Department of Science and Innovation for funding this project through the South African Quantum Technology Initiative (SAQuTI). We also acknowledge the Centre for High Performance Computing (CHPC) for granting us access to the Lengau cluster. JJB acknowledges support from U.S. Department of Energy, Office of Science, Quantum Systems Accelerator, from AFOSR, and from the DARPA ONISQ program. We thank A. Carter, M. Khan and K. Beloy for useful comments and discussions. Lastly, sincere thanks to Prof.~Hermann Uys for his contributions in developing the PS state protocol at its inception.  
\appendix
\section{Modified measurement operators}

In the absence of decoherence, the original protocol and its modification presented here produce identical post-measurement states after step 5, as both protocols are well-defined in the symmetric Dicke subspace. Ref.~\cite{alexander2020generating} introduces the original set of measurement operators defined by \begin{align}
\begin{split}
    \hat{A}_c := \sum_{m}&\sqrt{\text{Pr}\big(M\big|c\big)}\bigg|\frac{N}{2}, M \bigg\rangle \bigg\langle \frac{N}{2}, M\bigg|,
    \label{A1}
\end{split}
\end{align} 
where $M = \frac{N}{2}-m$ is the spin projection in the z-direction, and $m$ is the number of spin down qubits. $\text{Pr}(x|c)$ is given by Eq.~(\ref{Pr}), and the Dicke basis states by Eq.~(\ref{A2})
\begin{align}
\begin{split}
    \hspace{0.2cm}
    &\bigg|\frac{N}{2}, \frac{N}{2}-m\bigg\rangle \\&:= \frac{\sqrt{m!(N-m)!}}{\sqrt{N!}}\sum_{j_1<\cdot\cdot\cdot<j_m}|\uparrow \cdot\cdot\cdot\downarrow_{j_1}\cdot\cdot\cdot\downarrow_{j_m}\cdot\cdot\cdot\uparrow \rangle. \notag
\end{split}
\end{align}

In the ideal decoherence-free setting, the rotated squeezed coherent spin state generated during step 4 of the protocol, is well-defined in the subspace spanned by the Dicke basis states~\cite{alexander2020generating}, and we can therefore represent it as a linear combination thereof, 
\begin{align}
    |\psi\rangle = \text{exp}\bigg(i\frac{\pi}{2}\hat{J_x}\bigg) \hat{U}_{Sq}(\chi t)|CS\rangle =  \sum_{m}a_m\bigg|\frac{N}{2}, \frac{N}{2}-m\bigg\rangle,
    \label{A3}
\end{align}
where $a_m \in \mathbb{C}$. Both sets of measurement operators, as defined by Eqs.~(\ref{Ac}) and (\ref{A1}), lead to identical post-measurement states of the form
\begin{align}
    \hat{A}_c|\psi\rangle = \sum_{m}a_m\sqrt{\text{Pr}\bigg(\frac{N}{2}-m\Big|c\bigg)}\bigg|\frac{N}{2}, \frac{N}{2}-m\bigg\rangle.
    \label{A4}
\end{align}
Hence, the probability of obtaining outcome $c$ is identical and can be expressed as $\text{P}(c) = \text{Tr}[\hat{A}_{c}^{\dagger}\hat{A}_{c}\rho] = \langle \psi|\hat{A}_{c}^{\dagger}\hat{A}_{c}|\psi\rangle = \sum_{m}|a_m|^2\text{Pr}\big(N/2-m|c\big)$.
Thus, in the assumed decoherence-free setting, the measurement operators defined by Eqs.~(\ref{Ac}) and (\ref{A1}) are equivalent.
\section{Quantum trajectory method post-selection}
Consider the pre-measurement rotated squeezed state $\rho = \sum_ip_i|\psi_i\rangle\langle \psi_i|$, where $p_i = 1/\mathcal{N}$, obtained by employing the quantum trajectory method during the spin squeezing step of the PS state protocol. Measurement outcome $c$ yields the post-measurement state described by the mapping 
\begin{align}
\rho \mapsto \frac{\hat{A}_{c}\rho \hat{A}_{c}^{\dagger}}{\text{Tr}[\hat{A}_{c}\rho \hat{A}_{c}^{\dagger}]} &= \sum_i p_i \frac{\hat{A}_c |\psi_i\rangle\langle \psi_i|\hat{A}_c^{\dagger}}{\text{Tr}[\hat{A}_c^{\dagger}\hat{A}_{c}\rho]}  \\& = \sum_i \frac{\text{P}(i|c)\text{P}(c)}{\text{P}(c|i)} \frac{\hat{A}_c |\psi_i\rangle\langle \psi_i|\hat{A}_c^{\dagger}}{\text{Tr}[\hat{A}_c^{\dagger}\hat{A}_{c}\rho]} \\& = \sum_i \text{P}(i|c) \frac{\hat{A}_c |\psi_i\rangle\langle \psi_i|\hat{A}_c^{\dagger}}{\text{Tr}[\hat{A}_c^{\dagger}\hat{A}_{c}|\psi_i\rangle\langle \psi_i|]}\\& = \sum_i \text{P}(i|c) |\psi^c_i\rangle\langle \psi_i^c|,
\end{align}
with $\text{P}(i|c)$ and $\text{P}(c|i)$ denoting the conditional probabilities, and where 
\begin{align}
|\psi_i^c\rangle = \frac{\hat{A}_c|\psi_i\rangle}{\langle \psi_i|\hat{A}_c^{\dagger}\hat{A}_c|\psi_i\rangle^{1/2}}
\end{align}
denotes the post-measurement state of the $i$-th trajectory, given measurement outcome $c$.
\section{Measurements in the rotating reference frame}
Consider a pure state in the lab reference frame 
\begin{align}
|\psi_L\rangle = \hat{U}|\psi_R\rangle,
\label{o}
\end{align} 
where $\hat{U}(t) := \text{exp}(-i t \hat{H}/\hbar)$. The state $|\psi_R\rangle$ corresponds to the state $|\psi_L\rangle$ in the rotating reference frame, which is characterized by the unitary transformation $\hat{U}$. 

The measurement operator in the lab frame, denoted by $\hat{A}_{L}$, undergoes a unitary transformation defined by $\hat{U}(t)$, yielding the measurement operator in the rotating reference frame as \begin{align}
\hat{A}_{R} = \hat{U}^{\dagger}\hat{A}_L \hat{U}.
\label{AR}
\end{align}
Therefore, the act of measurement in the lab frame using $\hat{A}_{L}$ on $|\psi_L\rangle$ is equivalent to the act of measurement in the rotating frame using $\hat{A}_{R}$ on $|\psi_R\rangle$. Now consider the measurement operators given by Eq.~(\ref{Ac}), with the Hamiltonian $\hat{H} = \hbar \omega \hat{\sigma}^z$ (which is equal to $\hbar \omega \hat{\sigma}^+\hat{\sigma}^-$ up to a global phase factor in state). The unitary operator is given by 
\begin{align}
\hat{U}(\theta) &= \text{exp}\bigg(-i\frac{\theta}{2}\sum_j \hat{\sigma}^z_j\bigg) \\&
= \prod_{j}\hat{R}^{j}_{z}(\theta) \label{uniz},
\end{align}
where $\theta = 2 \omega t$. Consider the measurement operators of Eq.~(\ref{Ac}) having undergone the rotated reference frame transformation given by Eq.~(\ref{AR}).  The corresponding unitary operator given by Eq.~(\ref{uniz}) acts on the projector terms of the measurement operators as  
\begin{align}
\hat{U}^{\dagger}&|\underbrace{\big| \uparrow \cdots \downarrow_{j_1}\cdots \downarrow_{j_m} \cdots\uparrow \big\rangle}_\text{$N-m$ spin up} \notag \\& =   \prod_{j}\hat{R}^{j}_{z}(-\theta)\big| \uparrow \cdots \downarrow_{j_1}\cdots \downarrow_{j_m} \cdots\uparrow \big\rangle \\&
= e^{i \theta (N-m)/2}e^{-i\theta m/2}\big| \uparrow \cdots \downarrow_{j_1}\cdots \downarrow_{j_m} \cdots\uparrow \big\rangle,
\end{align}
and 
\begin{align}
\big\langle \uparrow \cdots \downarrow_{j_1}\cdots \downarrow_{j_m} \cdots\uparrow \big|\hat{U} \notag \\= \big\langle \uparrow \cdots \downarrow_{j_1}\cdots \downarrow_{j_m} \cdots\uparrow \big|e^{-i \theta (N-m)/2}e^{i\theta m/2}.
\end{align}
Given that the phases cancel in each term of the summation, it follows directly that $\hat{A}_{c,L} = \hat{A}_{c,R}$. This shows that the measurements given by Eq.~(\ref{Ac}) are equivalent in the lab and rotating reference frames. 
\\

\begin{figure*} [tbp]
\hspace*{0.cm}
\setlength{\tabcolsep}{0.15cm}
    \begin{table*}
        
        \centering
        \hspace*{-.28cm}
\renewcommand{\arraystretch}{2.0}
        \begin{tabular}{|c|c||c|c||c|c|}

\hline
    \rowcolor[gray]{0.89}[0.15cm]
        \multicolumn{6}{|c|}{\boldsymbol{$x= 0$},~~~\boldsymbol{$N = 4$},~~~ \boldsymbol{$\sigma^2 = 1.1,~~~\mathcal{F}_{\text{MSS}}=1.$}}\\
        \hline
        $\mathscr{R}(\mathcal{F})_{c \in [-2.5,1.5]}$  & \text{Pr}[$-2.5 \leq c \leq 1.5$] & $\mathscr{R}(\mathcal{F})_{c \in [-2.5,-0.5]}$ & \text{Pr}[$-2.5 \leq c \leq -0.5$] & $\mathscr{R}(\mathcal{F})_{c \in [-2.5,-1.5]}$ & \text{Pr}[$-2.5 \leq c \leq -1.5$] \\
        \hline
        $[0.21, 0.99]$ & $0.34$ & $[0.60, 0.99]$ & $0.02$ & $[0.90, 0.99]$ & $0.003$ \\
        \hline

    \rowcolor[gray]{0.89}[0.15cm]
        \multicolumn{6}{|c|}{\boldsymbol{$x = 1$},~~~\boldsymbol{$N = 4$},~~~ \boldsymbol{$ \sigma^2 = 1.1,~~~\mathcal{F}_{\text{MSS}}=0.79.$}}\\
        \hline
        $\mathscr{R}(\mathcal{F})_{c \in [-2.5,1.5]}$  & \text{Pr}[$-2.5 \leq c \leq 1.5$] & $\mathscr{R}(\mathcal{F})_{c \in [-2.5,-0.5]}$ & \text{Pr}[$-2.5 \leq c \leq -0.5$] & $\mathscr{R}(\mathcal{F})_{c \in [-2.5,-1.5]}$ & \text{Pr}[$-2.5 \leq c \leq -1.5$] \\
        \hline
        $[0.20, 0.87]$ & 0.35 & $[0.53, 0.87]$ & 0.02 & $[0.78, 0.87]$ & 0.004 \\
        \hline

    \rowcolor[gray]{0.89}[0.15cm]
        \multicolumn{6}{|c|}{\boldsymbol{$x = 0$},~~~\boldsymbol{$N = 6$},~~~ \boldsymbol{$\sigma^2 = 1.3,~~~\mathcal{F}_{\text{MSS}}=1.$}}\\
        \hline
        $\mathscr{R}(\mathcal{F})_{c \in [-2.5,1.5]}$  & \text{Pr}[$-2.5 \leq c \leq 1.5$] & $\mathscr{R}(\mathcal{F})_{c \in [-2.5,-0.5]}$ & \text{Pr}[$-2.5 \leq c \leq -0.5$] & $\mathscr{R}(\mathcal{F})_{c \in [-2.5,-1.5]}$ & \text{Pr}[$-2.5 \leq c \leq -1.5$] \\
        \hline
        $[0.19, 0.99]$ & $0.17$ & $[0.62, 0.99]$ & $0.01$ & $[0.88, 0.99]$ & $0.003$ \\
  \hline
      \rowcolor[gray]{0.89}[0.15cm]
        \multicolumn{6}{|c|}{\boldsymbol{$x= 1$},~~~\boldsymbol{$N = 6$},~~~ \boldsymbol{$\sigma^2 = 1.3,~~~\mathcal{F}_{\text{MSS}}=0.56.$}}\\
        \hline
        $\mathscr{R}(\mathcal{F})_{c \in [-2.5,1.5]}$  & \text{Pr}[$-2.5 \leq c \leq 1.5$] & $\mathscr{R}(\mathcal{F})_{c \in [-2.5,-0.5]}$ & \text{Pr}[$-2.5 \leq c \leq -0.5$] & $\mathscr{R}(\mathcal{F})_{c \in [-2.5,-1.5]}$ & \text{Pr}[$-2.5 \leq c \leq -1.5$] \\
        \hline
        $[0.17, 0.82]$ & $0.18$ & $[0.51, 0.82]$ & $0.02$ & $[0.71, 0.82]$ & $0.003$ \\
  \hline

      \rowcolor[gray]{0.89}[0.15cm]
        \multicolumn{6}{|c|}{\boldsymbol{$x = 0$},~~~\boldsymbol{$N = 8$},~~~ \boldsymbol{$\sigma^2 = 1.5,~~~\mathcal{F}_{\text{MSS}}=1.$}}\\
        \hline
        $\mathscr{R}(\mathcal{F})_{c \in [-2.5,1.5]}$  & \text{Pr}[$-2.5 \leq c \leq 1.5$] & $\mathscr{R}(\mathcal{F})_{c \in [-2.5,-0.5]}$ & \text{Pr}[$-2.5 \leq c \leq -0.5$] & $\mathscr{R}(\mathcal{F})_{c \in [-2.5,-1.5]}$ & \text{Pr}[$-2.5 \leq c \leq -1.5$] \\
        \hline
        $[0.22, 0.98]$ & 0.10 & $[0.66, 0.98]$ & 0.01 & $[0.88, 0.98]$ &  0.003 \\
  \hline
      \rowcolor[gray]{0.89}[0.15cm]
        \multicolumn{6}{|c|}{\boldsymbol{$x = 1$},~~~\boldsymbol{$N = 8$},~~~ \boldsymbol{$\sigma^2 = 1.5,~~~\mathcal{F}_{\text{MSS}}=0.33.$}}\\
        \hline
        $\mathscr{R}(\mathcal{F})_{c \in [-2.5,1.5]}$  & \text{Pr}[$-2.5 \leq c \leq 1.5$] & $\mathscr{R}(\mathcal{F})_{c \in [-2.5,-0.5]}$ & \text{Pr}[$-2.5 \leq c \leq -0.5$] & $\mathscr{R}(\mathcal{F})_{c \in [-2.5,-1.5]}$ & \text{Pr}[$-2.5 \leq c\leq -1.5$] \\
        \hline
        $[0.17,0.76]$ & 0.11 & $[0.50,0.76]$ & 0.01 & $[0.67,0.76]$ & 0.003 \\
  \hline
          \rowcolor[gray]{0.89}[0.15cm]
        \multicolumn{6}{|c|}{\boldsymbol{$x = 0$},~~~\boldsymbol{$N = 10$},~~~ \boldsymbol{$\sigma^2 = 1.6,~~~\mathcal{F}_{\text{MSS}} = 1.$}}\\
        \hline
        $\mathscr{R}(\mathcal{F})_{c \in [-2.5,1.5]}$  & \text{Pr}[$-2.5 \leq c \leq 1.5$] & $\mathscr{R}(\mathcal{F})_{c \in [-2.5,-0.5]}$ & \text{Pr}[$-2.5 \leq c \leq -0.5$] & $\mathscr{R}(\mathcal{F})_{c \in [-2.5,-1.5]}$ & \text{Pr}[$-2.5 \leq c \leq -1.5$] \\
        \hline
        $[0.28, 0.98]$ & 0.06 & $[0.74, 0.98]$ & 0.01 & $[0.91, 0.98]$ & 0.003\\
  \hline
        \rowcolor[gray]{0.89}[0.15cm]
        \multicolumn{6}{|c|}{\boldsymbol{$x = 1$},~~~\boldsymbol{$N = 10$},~~~ \boldsymbol{$\sigma^2 = 1.6,~~~\mathcal{F}_{\text{MSS}} = 0.17.$}}\\
        \hline
        $\mathscr{R}(\mathcal{F})_{c \in [-2.5,1.5]}$  & \text{Pr}[$-2.5 \leq c \leq 1.5$] & $\mathscr{R}(\mathcal{F})_{c \in [-2.5,-0.5]}$ & \text{Pr}[$-2.5 \leq c \leq -0.5$] & $\mathscr{R}(\mathcal{F})_{c \in [-2.5,-1.5]}$ & \text{Pr}[$-2.5 \leq c \leq -1.5$] \\
        \hline
        $[0.20,0.71]$ & 0.08 & $[0.52,0.71]$ & 0.01 & $[0.65,0.71]$ & 0.004 \\
  \hline
            \rowcolor[gray]{0.89}[0.15cm]
        \multicolumn{6}{|c|}{\boldsymbol{$x = 0$},~~~\boldsymbol{$N = 12$},~~~ \boldsymbol{$\sigma^2 = 1.8,~~~\mathcal{F}_{\text{MSS}} = 1.$}}\\
        \hline
        $\mathscr{R}(\mathcal{F})_{c \in [-2.5,1.5]}$  & \text{Pr}[$-2.5 \leq c \leq 1.5$] & $\mathscr{R}(\mathcal{F})_{c \in [-2.5,-0.5]}$ & \text{Pr}[$-2.5 \leq c \leq -0.5$] & $\mathscr{R}(\mathcal{F})_{c \in [-2.5,-1.5]}$ & \text{Pr}[$-2.5 \leq c \leq -1.5$] \\
        \hline
        $[0.34, 0.98]$ & 0.05 & $[0.76, 0.98]$ & 0.01 & $[0.92, 0.98]$ & 0.004\\
    \hline
        \end{tabular}
        \caption{The Kraus operator formalism given by Eq.~(\ref{Kr}), with $x \in \{0,1\}$, $dt = 1\times 10^{-7}$, and $N \in \{4, 6 , 8 , 10, 12\}$, for a comparative GHZ overlap fidelity study of the PS state and MSS. The included dephasing, amplitude damping and spontaneous excitation channels are characterized by Eqs.~(\ref{dec}) and (\ref{dec3}). We consider the range $\mathscr{R}(\mathcal{F})$ (the minimum and maximum $\mathcal{F}$) over chosen measurement intervals, as well as the comparative $\mathcal{F}_{\text{MSS}}$ values. It is worth emphasizing that for each considered value of $N$, the maximum fidelity $\mathcal{F}$ is obtained when $c=-2.5$. The probability of post-selecting from an interval is denoted by Pr$[\cdot]$. In Fig.~\ref{5q} and Fig.~\ref{QFIp}, we investigate the $N=12$ case with $x=1$ using the quantum trajectory method, focusing on just the optimal outcome of $c=-2.5$ due to the computational challenge of including a range as done for $N=4$ to $10$, in this table.} 
      
        \label{TAB}
    \end{table*}
\end{figure*}

\section{Experimental decay parameters}
Let us examine the scenario of a single-qubit system to provide context for Eq.~(\ref{Gam2}), which establishes a connection between the decay parameters ${\Gamma_{\text{dep}}, \Gamma_{\text{ad}}, \Gamma_{\text{se}}}$ and the Hamiltonian parameter $J$. In Ref.~\cite{foss2013nonequilibrium}, an alternative representation of the final term in the Lindblad master equation, as expressed in Eq.~(\ref{Lind}), can be denoted as 
\begin{align}
\sum_{j}\hat{L}_{j}\rho\hat{L}^{\dagger}_{j} = \frac{\mathcal{D}(\rho)}{\hbar},
\end{align} 
where 
\begin{align}
\mathcal{D}(\rho) := 2 \sum_{\text{all }\mathcal{J}}\mathcal{J}\rho\mathcal{J}^{\dagger},
\end{align}
and 
\begin{align}
\mathcal{J} \in \bigg\{\sqrt{\frac{\gamma_{\text{dep}}}{8}}\hat{\sigma}^{z}, \sqrt{\frac{\gamma_{\text{ad}}}{2}}\hat{\sigma}^{-},\sqrt{\frac{\gamma_{\text{se}}}{2}}\hat{\sigma}^{+} \bigg\}. 
\end{align}
It follows that 
\begin{align}
\mathcal{D}(\rho) = &\, \frac{1}{4}\gamma_{\text{dep}} \hat{\sigma}^{z}\rho \hat{\sigma}^{z} + \gamma_{\text{ad}}\hat{\sigma}^{-}\rho \hat{\sigma}^{-} \notag + \gamma_{\text{se}}\hat{\sigma}^{+}\rho \hat{\sigma}^{+}\\
& = \hbar\Gamma_{\text{dep}} \hat{\sigma}^{z}\rho \hat{\sigma}^{z} + \hbar\Gamma_{\text{ad}}\hat{\sigma}^{-}\rho \hat{\sigma}^{-} \notag + \hbar\Gamma_{\text{se}}\hat{\sigma}^{+}\rho \hat{\sigma}^{+},
\end{align}
where
\begin{align}
    \Gamma := \bigg\{\frac{1}{4}\frac{\gamma_{\text{dep}}}{\hbar}, \frac{\gamma_{\text{ad}}}{\hbar}, \frac{\gamma_{\text{se}}}{\hbar} \bigg\}.
\end{align}
Let $\gamma := \frac{1}{2}(\gamma_{\text{ad}}+\gamma_{\text{se}}+\gamma_{\text{dep}})$; typical experimental numbers are quoted as $\gamma = 0.05 \hbar J$ and $\gamma_{\text{dep}} = 8\gamma_{\text{ad}} = 8\gamma_{\text{se}}$ (see Refs.~\cite{foss2013nonequilibrium, bohnet2016quantum}). Note that our $J$ does not include $\hbar$. For the dephasing channel it follows that 
\begin{align}
0.05\hbar J = &\, \frac{1}{2}\bigg(\frac{\gamma_{\text{dep}}}{8}+\frac{\gamma_{\text{dep}}}{8}+ \gamma_{\text{dep}}\bigg)\\
& = \frac{5}{8}\gamma_{\text{dep}}        \\
& = \frac{5}{2} \hbar \Gamma_{\text{dep}}.         
\end{align}
This implies that 
\begin{align}
    J = 50\Gamma_{\text{dep}}.
\end{align}
Similar arguments apply when considering the amplitude damping and spontaneous excitation channels, leading to Eq.~(\ref{Gam2}).
\section{Post-selection analysis}
In Tab.~\ref{TAB} we provide a comparative analysis of the GHZ overlap fidelity with the PS and MSS states in the ideal ($x=0$) and decoherence ($x=1$) cases for increasing qubit number $N$, together with the post-selection success probability for generating the PS state over a range of $c$ values.


\begin{thebibliography}{9}

\bibitem{schlosshauer2019quantum} M. Schlosshauer, ``Quantum decoherence”, {\it{Physics Reports}}, {\bf{831}}, 1–57, (2019).

\bibitem{nielsen2002quantum} M. A. Nielsen and I. L. Chuang, {\textit{Quantum Computation and Quantum Information: 10th Anniversary Edition}}, (Cambridge University Press, Cambridge, 2010).

\bibitem{horodecki2009quantum} R. Horodecki, P Horodecki, M. Horodecki, and K. Horodecki,  ``Quantum entanglement", {\it{Reviews of Modern Physics}}, {\bf{81}}, 865, (2009).
\bibitem{greenberger1989going} D. M. Greenberger, M. A. Horne, and A. Zeilinger, Going beyond Bell’s theorem, in {\it{Bell’s theorem, quantum theory and conceptions of the universe}}, (Springer, Dordrecht, 1989), pp. 69-72.

\bibitem{zheng2000efficient}  S.-B. Zheng and G.-C. Guo, ``Efficient scheme for two-
atom entanglement and quantum information processing in cavity QED”, {\it{Physical Review Letters}}, {\bf{85}}, 2392, (2000). 

\bibitem{schaffry2010quantum}  M. Schaffry, E. M. Gauger, J. J. Morton, J. Fitzsimons,
S. C. Benjamin, and B. W. Lovett, ``Quantum metrology  with molecular ensembles”, {\it{Physical Review A}}, {\bf{82}}, 042114, (2010).

\bibitem{pezze2018quantum} L. Pezz{\`e}, A. Smerzi, M. K. Oberthaler, R. Schmied, and P. Treutlein, ``Quantum metrology with nonclassical states of atomic ensembles", {\it{Reviews of Modern Physics}}, {\bf{90}}, 035005, (2018).

\bibitem{broadbent2009ghz}  A. Broadbent, P.-R. Chouha, and A. Tapp, ``The GHZ state in secret sharing and entanglement simulation”, {\it{In
2009 Third International Conference on Quantum, Nano and Micro Technologies}}, 59–62, (IEEE, 2009).

\bibitem{zhu2018semi} K.-N. Zhu, N.-R. Zhou, Y.-Q. Wang, and X.-J. Wen,
``Semi-quantum key distribution protocols with GHZ states”, {\it{International Journal of Theoretical Physics}}, {\bf{57}}, 3621-3631, (2018).

\bibitem{gao2005deterministic} T. Gao, F.-L. Yan, and Z.-X. Wang, ``Deterministic secure direct communication using GHZ states and swapping quantum entanglement”, {\it{Journal of Physics A: Mathematical and General}}, {\bf{38}}, 5761, (2005).

\bibitem{jin2006three}  X.-R. Jin, X. Ji, Y.-Q. Zhang, S. Zhang, S.-K. Hong,
K.-H. Yeon, and C.-I. Um, ``Three-party quantum secure direct communication based on GHZ states,” {\it{Physics Letters A}}, {\bf{354}}, 67-70, (2006).
\bibitem{song2019generation} C. Song, K. Xu, H. Li, Y. R. Zhang, X. Zhang, W. Liu, Q. Guo, Z. Wang, W. Ren, J. Hao {\it{et al}}., Generation of multicomponent atomic Schrödinger cat states of up to 20 qubits, {\it{Science}}, {\bf{365}}, 574–577, (2019).

\bibitem{leibfried2005creation} D. Leibfried, E. Knill, S. Seidelin, J. Britton, R. B. Blakestad, J. Chiaverini, D. B. Hume, W. M. Itano, J. D. Jost, C. Langer {\it{et al}}., Creation of a six-atom ‘Schr{\"o}dinger cat’ state, {\it{Nature}}, {\bf{438}}, 639--642, (2005).
\bibitem{monz201114} T. Monz, P. Schindler, J. T. Barreiro, M. Chwalla, D. Nigg, W. A. Coish, M. Harlander, W. Hänsel, M. Hennrich, and R. Blatt, 14-qubit entanglement: Creation and coherence, {\it{Phys. Rev. Lett.}}, {\bf{106}}, 130506, (2011).

\bibitem{omran2019generation} A. Omran, H. Levine, A. Keesling, G. Semeghini, T. T. Wang, S. Ebadi, 
H. Bernien, A. S. Zibrov, H. Pichler, S. Choi {\textit{et al}}.,  Generation and manipulation of Schrödinger cat states in Rydberg atom arrays, {\it{Science}}, {\bf{365}}, 570-574, (2019).

\bibitem{plodzien2020producing} M. Płodzień, M. Kościelski, E. Witkowska, and A. Sinatra, ``Producing and storing spin-squeezed states and Greenberger-Horne-Zeilinger states in a one-dimensional optical lattice", {\it{Physical Review A}}, {\bf{102}}, 013328, (2020).

\bibitem{xia2016generating} K. Xia and J. Twamley, ``Generating spin squeezing
states and Greenberger-Horne-Zeilinger entanglement using a hybrid phonon-spin ensemble in diamond”, {\it{Physical
Review B}}, {\bf{94}}, 205118, (2016).

\bibitem{gross2012spin}  C. Gross, ``Spin squeezing, entanglement and quantum metrology with Bose--Einstein condensates”, {\it{Journal of Physics B: Atomic, Molecular and Optical Physics}}, {\bf{45}}, 103001, (2012).

\bibitem{foss2013nonequilibrium}  M. Foss-Feig, K. R. Hazzard, J. J. Bollinger, and A. M. Rey, ``Nonequilibrium dynamics of arbitrary-range Ising models with decoherence: An exact analytic solution”, {\it{Physical Review A}}, {\bf{87}}, 042101, (2013).

\bibitem{helmer2009measurement} F. Helmer and F. Marquardt, ``Measurement-based synthesis of multiqubit entangled states in superconducting cavity QED”, {\it{Physical
Review A}}, {\bf{79}}, 052328, (2009).

\bibitem{bishop2009proposal} L. S. Bishop, L. Tornberg, D. Price, E. Ginossar, A. Nunnenkamp,
A. A. Houck, J. M. Gambetta, J. Koch, G. Johansson, S. M. Girvin \textit{et al}., Proposal for generating and detecting multi-qubit GHZ states in circuit QED, {\it{New Journal of Physics}}, {\bf{11}}, 073040, (2009).
\bibitem{toth2005entanglement} G. T{\'o}th and O. G{\"u}hne, ``Entanglement detection in the stabilizer”, {\it{Physical Review A}}, {\bf{72}}, 022340,
(2005).

\bibitem{alexander2020generating} B. Alexander, J. J. Bollinger, and H. Uys, ``Generating Greenberger-Horne-Zeilinger states with squeezing and
postselection”, {\it{Physical Review A}}, {\bf{101}}, 062303,
(2020).

\bibitem{schleier2010squeezing} M. H. Schleier-Smith, I. D. Leroux, and V. Vuleti{\'c}, ``Squeezing the collective spin of a dilute atomic ensemble by cavity feedback", {\it{Physical Review A}}, {\bf{81}}, 021804, (2010).

\bibitem{zhang2012collective} H. Zhang, R.McConnell, S. {\'C}uk, Q. Lin,M. H. Schleier-Smith,
I. D. Leroux, and V. Vuleti{\'c}, ``Collective state measurement of mesoscopic ensembles with single-atom resolution", {\it{Physical Review Letters}}, {\bf{109}}, 133603, (2012).

\bibitem{hosten2016measurement} O. Hosten, N. J. Engelsen, R. Krishnakumar, and M. A. Kasevich, ``Measurement noise 100 times lower than the
quantum-projection limit using entangled atoms”, {\it{Nature}}, {\bf{529}}, 505, (2016).

\bibitem{chen2011conditional} Z. Chen, J. G. Bohnet, S. R. Sankar, J. Dai, and J. K. Thompson, ``Conditional spin squeezing of a large ensemble via the vacuum Rabi splitting”, {\it{Physical Review Letters}}, {\bf{106}}, 133601, (2011).

\bibitem{carlo2004simulating} G. G. Carlo, G. Benenti, G. Casati, and C. Mejia-Monasterio, ``Simulating noisy quantum protocols with quantum trajectories", {\it{Physical Review A}}, {\bf{69}}, 062317, (2004).

\bibitem{pezze2019heralded} L. Pezz{\`e}, M. Gessner, P. Feldmann, C. Klempt, L. Santos, and A. Smerzi, ``Heralded generation of macroscopic superposition states in a spinor Bose-Einstein condensate", {\it{Physical Review Letters}}, {\bf{123}}, 260403, (2019).

\bibitem{derpanis2008bhattacharyya} K. G. Derpanis, ``The bhattacharyya measure”, {\it{Mendeley Computer}}, {\bf{1}}, 1990-1992, (2008).

\bibitem{bourennane2004experimental} M. Bourennane, M. Eibl, C. Kurtsiefer, S. Gaertner, H. Weinfurter, O. Gühne, P. Hyllus, D. Bruß, M. Lewenstein, and A. Sanpera, Experimental detection of multipartite entanglement using witness operators, {\it{Phys. Rev. Lett.}}, {\bf{92}}, 087902, (2004).
\bibitem{zhu2022experimental} G. Zhu, C. Zhang, K. Wang, L. Xiao, and P. Xue, ``Experimental witnessing for entangled states with limited local measurements”, {\it{Photonics Research}}, {\bf{10}}, 2047--2055, (2022).

\bibitem{petz2011introduction}
D. Petz and C. Ghinea, {\sl Quantum probability and related topics - Proceedings of the 30th Conference}, (World Scientific, Singapore, 2011).

\bibitem{paris2009quantum} M. G. Paris, ``Quantum estimation for quantum technology", {\it{International Journal of Quantum Information}}, {\bf{7}}, 125–137, (2009).

\bibitem{helstrom1969quantum} C. W. Helstrom, ``Quantum detection and estimation theory", {\it{Journal of Statistical Physics}}, {\bf{1}}, 231–252, (1969).

\bibitem{braunstein1994statistical} S. L. Braunstein and C. M. Caves, ``Statistical distance and the geometry of quantum states", {\it{Physical Review Letters}}, {\bf{72}}, 3439, (1994).

\bibitem{lu2015robust} X. M. Lu, S. Yu, and C. H. Oh, Robust quantum metrological schemes based on protection of quantum Fisher information, {\it{Nature Communications}}, {\bf{6}}, 7282, (2015).

\bibitem{arvidsson2020quantum} D. R. Arvidsson-Shukur, N. Yunger Halpern, H. V. Lepage, A. A. Lasek,  C. H. Barnes, and  S. Lloyd, ``Quantum advantage in postselected metrology", {\it{Nature Communications}}, {\bf{11}}, 3775, (2020).

\bibitem{toth2012multipartite} G. T{\'o}th, ``Multipartite entanglement and high-precision metrology", {\it{Physical Review A}}, {\bf{85}}, 022322, (2012).

\bibitem{toth2014quantum} G. T{\'o}th and I. Apellaniz, ``Quantum metrology from a quantum information science perspective", {\it{Journal of Physics A: Mathematical and Theoretical}}, {\bf{47}}, 424006, (2014). 

\bibitem{dicke1954coherence} R. H. Dicke, ``Coherence in spontaneous radiation processes”, {\it{Physical Review}}, {\bf{93}}, 99, (1954).

\bibitem{gardiner1992wave} C. W. Gardiner, A. S. Parkins, and P. Zoller, ``Wave-function quantum stochastic differential equations and
quantum-jump simulation methods”, {\it{Physical Review A}}, {\bf{46}}, 4363, (1992).

\bibitem{breuer1995stochastic} H.-P. Breuer and F. Petruccione, ``Stochastic dynamics of quantum jumps”, {\it{Physical Review E}}, {\bf{52}}, 428, (1995).

\bibitem{kitagawa1993squeezed} M. Kitagawa and M. Ueda, ``Squeezed spin states", {\it{Physical Review A}}, {\bf{47}}, 5138, (1993).

\bibitem{uys2012toward} H. Uys, M. Biercuk, J. Britton, and J. J. Bollinger, ``Toward spin squeezing with trapped ions”, {\it{AIP Conference Proceedings}}, {\bf{1469}}, 108–121, (2012).

\bibitem{husimi1940some}  K. Husimi, ``Some formal properties of the density matrix”, {\it{Proceedings of the Physico - Mathematical Society of
Japan. 3rd Series}}, {\bf{22}}, 264–314, (1940).

\bibitem{jacobs2014quantum} K. Jacobs, {\it{Quantum measurement theory and its applications}}, (Cambridge University Press, 2014).

\bibitem{metcalf2003laser} H. J. Metcalf and P. van der Straten, ``Laser cooling and trapping of atoms”, {\it{JOSA B}}, {\bf{20}}, 887–908,
(2003).


\bibitem{bohnet2016quantum} J. G. Bohnet, B. C. Sawyer, J. W. Britton, M. L. Wall, A. M. Rey, M. Foss-Feig, and J. J. Bollinger, ``Quantum spin dynamics and entanglement generation with hundreds of trapped ions”, {\it{Science}}, {\bf{352}}, 1297–-1301, (2016).

\bibitem{britton2012engineered} J. W. Britton, B. C. Sawyer, A. C. Keith, C.-C. J. Wang, J. K. Freericks, H. Uys, M. J. Biercuk, and J. J. Bollinger, ``Engineered two-dimensional Ising interactions in a trapped-ion quantum simulator with hundreds of spins”, {\it{Nature}}, {\bf{484}}, 489–492, (2012).

\bibitem{dalibard1992wave} J. Dalibard, Y. Castin, and K. Mølmer, ``Wave-function approach to dissipative processes in quantum optics”,
{\it{Physical Review Letters}}, {\bf{68}}, 580, (1992). 
\bibitem{blogpost} {\sl How Large Quantum Systems Can QuTiP Handle?}, Blog Archive,~(2012). Accessed on 1/7/2023. Accessible via \url{https://qutip.blogspot.com/2012/07/how-large-quantum-systems-can-qutip.html}. 

\bibitem{johansson2012qutip} J. R. Johansson, P. D. Nation, and F. Nori, ``QuTiP: An open-source Python framework for the dynamics of open quantum systems”, {\it{Comput. Phys. Commun.}}, {\bf{183}}, 1760--1772, (2012).

\bibitem{johansson2012qutip2} J. R. Johansson, P. D. Nation, and F. Nori, QuTiP 2: A Python framework for the dynamics of open quantum systems, {\it{Comput. Phys. Commun.}}, {\bf{184}}, 1234, (2013).

\bibitem{suzuki2021qulacs} Y. Suzuki, Y. Kawase, Y. Masumura, Y. Hiraga, M. Nakadai, J. Chen, K. M. Nakanishi, K. Mitarai, R. Imai, S. Tamiya \textit{et al}., Qulacs: a fast and versatile quantum circuit simulator for research purpose, {\it{Quantum}}, {\bf{5}}, 559, (2021).


\bibitem{zhang2023tensorcircuit} S. X. Zhang, J. Allcock, Z. Q. Wan, S. Liu, J. Sun, H. Yu, X. H Yang, J. Qiu, Z. Ye, Y. Q. Chen \textit{et al}., Tensorcircuit: a quantum software framework for the NISQ era, {\it{Quantum}}, {\bf{7}}, 912, (2023).

\bibitem{efthymiou2022quantum} S. Efthymiou, M. Lazzarin, A. Pasquale, and S. Carrazza, ``Quantum simulation with just-in-time compilation", 
{\textit{Quantum}}, {\textbf{6}}, 814, (2022).

\bibitem{CUNVid} H. Bayraktar, A. Charara, D. Clark, S. Cohen, T. Costa, Y. L. L. Fang, Y. Gao, J. Guan,
J. Gunnels, A. Haidar \textit{et al}., cuQuantum SDK: A high-performance library for accelerating quantum science, \textit{In 2023 IEEE International Conference on Quantum Computing and Engineering (QCE)}, (IEEE, Bellevue, USA, 2023), pp. 1050-1061.
\bibitem{manzano2020short} D. Manzano, ``A short introduction to the Lindblad master equation”, {\it{AIP Advances}}, {\bf{10}}, 025106, (2020).

\bibitem{breuer2002theory} H.-P. Breuer and F. Petruccione, {\it{The theory of open quantum systems}}, (Oxford University Press, New York, 2002).
\bibitem{kraus1983states} K.  Kraus, A. B{\"o}hm, J. D. Dollard, and W. H. Wootters, {\it{States, Effects, and Operations Fundamental Notions of Quantum Theory: Lectures in Mathematical Physics at the University of Texas at Austin}}, (Springer, 1983).

\bibitem{biercuk2009high} M. J. Biercuk, H. Uys, A. P. Vandevender, N. Shiga, W. M. Itano, and J. J. Bollinger, ``High-fidelity quantum control using ion crystals in a Penning trap”, {\it{Quantum Information \& Computation}}, {\bf{9}}, 920-949, (2009).

\bibitem{porras2004effective} D. Porras and J. I. Cirac, ``Effective quantum spin systems with trapped ions”, {\it{Physical Review Letters}}, {\bf{92}}, 207901, (2004).

\bibitem{jurcevic2014quasiparticle} P. Jurcevic, B. P. Lanyon, P. Hauke, C. Hempel, P. Zoller, R. Blatt, and C. F. Roos, ``Quasiparticle engineering and entanglement propagation in a quantum many-body system”, {\it{Nature}}, {\bf{511}}, 202-205, (2014).

\bibitem{elliott1970ising} R. J. Elliott, P. Pfeuty, and C. Wood, ``Ising model with a transverse field”, {\it{Physical Review Letters}}, {\bf{25}}, 443, (1970).

\bibitem{preskill2015lecture} J. Preskill, \textit{Lecture notes for ph219/cs219: Quantum information}, (2015). Accessible via \url{http://theory.caltech.edu/~preskill/ph219/chap3_15.pdf}.

\bibitem{wolfowicz2016pulse} G. Wolfowicz and J. J. Morton, ``Pulse techniques for quantum information processing", {\it{EMagRes}}, {\bf{5}}, 1515--1528, (2016).

\bibitem{bardin2021microwaves} J. C. Bardin, D. H. Slichter, and D. J. Reilly, ``Microwaves in quantum computing", {\it{IEEE Journal of Microwaves}}, {\bf{1}}, 403--427, (2021).

\bibitem{daley2014quantum} A. J. Daley, ``Quantum trajectories and open many-body quantum systems", {\it{Advances in Physics}}, {\bf{63}}, 77-149, (2014).


\end{thebibliography}
\end{document}